\documentclass[twocolumn]{aastex631}

\usepackage{amsmath}
\usepackage{subfigure}
\usepackage{color}
\usepackage{epsfig}
\usepackage{float}
\usepackage{bm}
\usepackage{multirow}
\usepackage{graphicx}
\usepackage{booktabs}

\begin{document}

\title{A Roadmap for Transient Hunters: Mapping Stellar Mass and Star Formation Rate Anisotropies in the Local Universe}

\correspondingauthor{Yuan-Pei Yang (ypyang@ynu.edu.cn)}

\author[0000-0001-8278-2955]{Ye-Hao Cheng}
\affiliation{South-Western Institute for Astronomy Research, Yunnan Key Laboratory of Survey Science, Yunnan University, Kunming, Yunnan 650504, People’s Republic of China}

\author[0000-0001-6374-8313]{Yuan-Pei Yang}
\affiliation{South-Western Institute for Astronomy Research, Yunnan Key Laboratory of Survey Science, Yunnan University, Kunming, Yunnan 650504, People’s Republic of China}

\author[0000-0001-5931-2381]{Ye Li}
\affiliation{Purple Mountain Observatory, Chinese Academy of Sciences, Nanjing, Jiangsu 210023, People’s Republic of China}
\affiliation{State Key Laboratory of Radio Astronomy and Technology, Purple Mountain Observatory, Chinese Academy of Sciences, 10 Yuanhua Road, Nanjing 210023, China}

\begin{abstract}
Over the past few decades, an increasing number of transients in nearby galaxies have been discovered through various survey projects. Unlike astrophysical phenomena at cosmological distances, transients in the local universe exhibit a pronounced anisotropy in their sky distribution. Consequently, adopting an appropriate survey strategy is essential to improve the efficiency of transient searches in the local universe. In this work, we utilized a large galaxy catalog to map the sky distributions of stellar mass and star formation rate (SFR) across different luminosity distance thresholds and angular resolutions of the grid on the celestial sphere. These maps can further serve to characterize the anisotropic spatial distribution of nearby extragalactic transients. For different angular resolutions of the celestial sphere, we find that the sky distributions of stellar mass of galaxies are similar to those of the SFR in the main anisotropic structures. As the luminosity distance threshold increases, the anisotropic structures of the sky distributions become more isotropic. We calculate the angular power spectra and fluctuations of the sky distribution of stellar mass and SFR at a given angular resolution and find that the angular power spectra and fluctuations decrease rapidly as the luminosity distance threshold increases. Finally, by qualitatively comparing the sky distribution of core-collapse supernovae with our SFR sky distribution, we find that the two exhibit consistent patterns in several prominent structures. The mapped sky distributions of stellar mass and SFR can serve as valuable references for future surveys in searching for extragalactic transients.
\end{abstract}


\section{Introduction} \label{sec:intro}

With advances in observational techniques, the detection capabilities of telescopes have greatly improved, leading to the discovery of numerous cosmic transients, e.g., gamma-ray bursts (GRBs) \citep{1973ApJ_GRB,1993ApJ...413L.101K} and their multiwavelength afterglows \citep{1997Nature_afterglow1,1997Nature_afterglow2}, supernovae (SNe) \citep{1934PNAS...20..254B,1938ApJ....88..529Z} , 
fast radio bursts (FRBs) \citep{2007Sci...318..777L,2013Sci...341...53T}, 
kilonovae \citep{2017Sci...358.1556C,2017ApJ...848L..16S,2017ApJ...848L..24V}, tidal disruption events (TDEs) \citep{1996A&A...309L..35B,1999A&A...343..775K}, fast blue optical transients (FBOTs) \citep{2014ApJ...794...23D,2018ApJ...865L...3P} and magnetar giant flare \citep{1999Natur.397...41H,2005Natur.434.1098H}, and so on. The discoveries of cosmic transients reveal the extreme astrophysical processes and provide some important probes for studying cosmology and fundamental physics. 

Among these transients, some of them (e.g., SNe, GRBs and FRBs) were discovered through various survey programs, while others (e.g., afterglows and kilonovae associated with GRBs) were more identified via Target of Opportunity (ToO) observations.
ToO observations are well suited for detecting phenomena associated with a triggering signal, because the trigger provides the transient’s localization and explosion time \citep{2021ApJ...918...63A}. 
For example, GRB triggers led to the discovery of multi-wavelength afterglows, which occur after the prompt sub-MeV emission phase \citep{1997Nature_afterglow1,1997Nature_afterglow2,1997ApJ...475..224F}. Furthermore, the first kilonova, AT2017gfo, was directly discovered by optical telescopes following the gravitational wave and short GRB signals from a binary neutron star merger in 2017 \citep{2017ApJGWevent1,2017ApJGWevent2,2017ApJGWevent3}.

Transients that lack high-energy triggers are more often discovered through survey programs. Transient survey program can generally be further divided into ``targeted surveys'' and ``un-targeted surveys''.
Historically, many discoveries of nearby transients were driven by the targeted surveys (e.g., SNe). Notable examples include the Lick Observatory Supernova Search (LOSS, \citealt{2000AIPC..522..103L, 2001ASPC..246..121F}) and the CHilean Automatic Supernova sEarch (CHASE, \citealt{2009AIPC.1111..551P, 2012MmSAI..83..388H}). These targeted surveys operated by monitoring pre-selected lists of prominent, massive, and luminous nearby galaxies. While these targeted surveys were highly successful in compiling early local SN catalogs, determining nearby SN rates \citep{2011MNRAS.412.1473L}, and anchoring the low-redshift Hubble diagram using Type Ia SN samples \citep{2010ApJS..190..418G, 2013MNRAS.433.2240G}, they inherently introduced significant observational biases into the host-galaxy demographics. This is because their target selection heavily favored luminous, high-mass, and metal-rich galaxies \citep{2010ApJ...721..777A, 2012ApJ...745...70P}. Although efforts were made to include faint galaxies, the low-luminosity end of their galaxy samples remained highly incomplete \citep{2011MNRAS.412.1419L}. Consequently, targeted surveys likely underestimated the number of transients occurring in low-surface-brightness galaxies, low-luminosity metal-poor dwarfs, and intergalactic environments. Therefore, relying solely on historical, targeted local catalogs can skew the scaling relations between host-galaxy populations and transients, and miss specific classes of transient events that preferentially reside in dwarf galaxies \citep{2012ApJ...745...70P}, such as superluminous supernovae (SLSNe, \citealt{2011Natur.474..487Q}). Furthermore, targeted surveys also suffered from spatial biases, as they typically neglected the outskirts of galaxies and intergalactic regions. Such spatial limitations could lead to the omission of events like compact object mergers and Calcium-rich gap transients \citep{2012PASA...29..482K,2017ApJ...836...60L,2020ApJ...905...58D}.

In contrast, modern untargeted wide-field surveys, such as Asteroid Terrestrial-impact Last Alert System (ATLAS, \citealt{2018PASP..130f4505T}), Zwicky Transient Facility (ZTF, \citealt{2019PASP..131a8002B}), and the Vera C. Rubin Observatory's Legacy Survey of Space and Time (LSST) \citep{2019ApJ...873..111I}, scan the sky regardless of known galaxy positions. This blind scanning strategy largely mitigates the aforementioned biases regarding host-galaxy mass, luminosity, and progenitor origins, thereby revealing the unbiased distribution of host galaxies and the true diversity of transients \citep{2020ApJ...904...35P}.

While historical targeted surveys (typically with small FoVs) focus on pre-selected massive galaxies, and modern wide-field surveys (with large FoVs) systematically and blindly scan the entire sky, a large fraction of astronomical facilities operate in the intermediate regime of medium FoVs (e.g., several square degrees), where optimized survey strategies are still lacking. For these medium-FoV telescopes, survey design faces a critical dilemma: blind scanning the entire sky is highly inefficient due to the vast empty spaces and low galaxy densities, which wastes precious telescope time and disrupts the required observational cadence; conversely, strictly targeting only the most luminous galaxies reintroduces severe observational biases and leads to the omission of transients in low-mass or dwarf hosts; moreover, even within a single pointing, the field of view often contains numerous galaxies, making such strict selection both impractical and inefficient.

Therefore, considering the strict constraints of FoV, cadence, and limited observing time, there is a pressing need for an optimized pointing strategy.
This is precisely the primary motivation of this work. 
By mapping the sky distributions of event rates at various angular resolutions, specifically matched to the FoVs of different telescopes, we provide a ``roadmap'' for transient hunters. Observers can consult these maps to strategically allocate their limited telescope time, prioritizing regions with the highest event-rate densities. This targeted approach ensures that every pointing, constrained by the telescope's specific FoV and cadence, yields the highest possible probability of discovering new transients.

In this work, we construct the sky distributions of event-rate of transients using the most complete galaxy catalog currently available. These distributions can guide the design of future untargeted survey strategies for transient discovery, especially for medium-FoV telescopes.
Since the universe is isotropic on large scales, the sky distributions of event-rate show significant anisotropy only within the local universe ($\lesssim 200~{\rm Mpc}$). As a result, these distributions are most effective for guiding transient searches in the local universe. 
Furthermore, the transients that dominate the local universe differ in certain respects from those typically observed at cosmological distances, making their detection both scientifically intriguing and timely: 
1) In observation, there exists a substantial gap between the luminosity of the brightest novae, with absolute magnitudes around $\sim -10$ mag, and that of sub-luminous supernovae, with absolute magnitudes around $\sim -16$ mag. Transients within this luminosity range are most likely to be detected in the local universe. 2) Some multi-messenger phenomena, such as gravitational waves, cosmic rays, and astrophysical neutrinos, are similarly constrained to the local universe due to various physical effects or instrumental sensitivities. Consequently, survey programs targeting transients in the local universe play a crucial role in multi-messenger astronomy.
3) Young astrophysical objects, which are relatively rare, are more readily observed in the local universe, where their detection is not hindered by distance or sensitivity limitations. Studying these nearby rare objects provides valuable insights into early stellar evolution and transient phenomena.

The paper is organized as follows. In Section \ref{sec:method}, we introduce the classification of transients, the galaxy sample used in this study and describe the methodology for predicting SFR based on machine learning and computing the sky distributions of stellar mass and SFR. In Section \ref{sec:results}, we present the resulting sky distributions of stellar mass and SFR. In Section \ref{sec:Discussion}, we provide a discussion of the results. The main conclusions are summarized in Section \ref{sec:Conclusion}.

\section{Sky Distributions of Stellar Mass and Star Formation Rate} \label{sec:method}
In this section, we describe the classification of transients, the methodology for predicting SFR based on machine learning and computing the sky distributions of galaxy stellar mass and SFR, including both the selection of the galaxy catalog and the details of the calculation procedure. 
Due to the finite sensitivity of telescopes, transient observations are limited to sources within a given luminosity distance. Although transient events appear approximately isotropic on cosmological scales, their distributions become highly inhomogeneous in the local universe. To account for these distance-dependent variations, we compute the sky distributions of stellar mass and SFR at multiple luminosity distance thresholds. This approach allows us to quantify how the underlying galaxy distribution influences transient event rates and to identify the most favorable regions of the sky for observations.

\subsection{Classification of Transients}
Most transients can be broadly classified into two categories. The first category comprises transients originating from old stellar populations, which are linked to the stellar mass distribution in the universe. 
These events may originate from low-mass stellar systems, such as from compact binary mergers, including short gamma-ray bursts, kilonovae, and Type Ia supernovae produced by binary white dwarf mergers.
For example, gravitational-wave emission drives the merger of compact binaries long after the formation of their progenitors. The multi-wavelength transients produced by these merger events often occur at large offsets from the centers of their host galaxies, or are in some host galaxies with older stellar populations \citep{2017ApJ...848L..22B,2017ApJ...848L..23F,2017ApJ...848L..28L,2017ApJ...848L..30P,2017ApJ...849L..16I,2017ApJ...849L..34P}.

The second category comprises transients originating from young stellar populations, which are linked to the star formation rate (SFR) distribution in the universe. These transients typically result from the deaths of massive stars. Due to the relatively short lifetimes of massive stars, such transients are generally found in regions of active star formation. The most representative examples include core-collapse supernovae (CCSNe, e.g., SN Ib, SN Ic, SN II, and SLSNe) and long gamma-ray bursts \citep{2006ARA&ASN}.
In addition to the catastrophic events mentioned above, phenomena associated with young neutron stars also typically originate in star-forming regions for the same reason. For example, magnetars could produce soft gamma repeaters \citep{1979SvAL....5..343M,1979Natur.282..587M,1981Ap&SS..75...47M}, giant flares \citep{1980ApJ...237L...7E,1999Natur.397...41H,2005Natur.434.1098H,2005ApJ...624L.105M,2007ApJ...661..458B,2022arXiv220507670Z} and fast radio bursts \citep{2020NatureFRB1,2020NatureFRB2} before their magnetic field significantly decays. Since the decay timescale is of the order of $\sim(10^3-10^5)$ years, these transients are usually associated with the SFR distribution in the universe.

However, it is important to emphasize that the dependence of transient rates on galaxy properties cannot be reduced to a simple dichotomy between stellar mass and SFR, as both quantities may jointly contribute to the observed transient rates. Different transients have different delay times between progenitor formation and explosion. It is well established that the relationship between transient rates, galaxy stellar mass, and SFR is fundamentally governed by the delay time distribution (DTD) of the progenitors \citep{2006MNRAS.370..773M,2012PASA...29..447M}. For transients with negligible delay times, such as CCSNe, the progenitor stars are massive and extremely short-lived compared to the cosmic timescale. 
As classically demonstrated, short-lived massive stars approximately trace the ongoing star formation activity \citep{2012ARA&A..50..531K}.
Consequently, the rates of these prompt transients are directly proportional to the current SFR and are predominantly found in late-type galaxies.

In contrast, for transients with non-negligible delay times, such as SN Ia, the occurrence rate reflects the integrated star formation history of the host galaxies. Extensive observational constraints on SN Ia progenitors \citep{2008PASJ...60.1327T,2014ARA&A..52..107M} have demonstrated that the DTD of SNe Ia follows a continuous power-law decay, spanning from very short delays ($\sim$ tens of millions of years) to extremely long delays ($\sim$ 10 billion years). This indicates that such transients do not explode at a fixed time after star formation, but rather occur across the long evolutionary history of a galaxy.
Therefore, it is physically inaccurate to assert that the event rates of long-delay transients depend solely on galaxy stellar mass. Instead, they are best described by the highly successful ``two-component model" (or ``A+B" model), which is given by
\begin{equation}
    \frac{{\rm  Rate}(t)}{10^{-2}~{\rm yr^{-1}galaxy^{-1}}}=A\left(\frac{M_{\star}(t)}{10^{10}~\rm M_{\odot}}\right) + B\left(\frac{{\rm SFR}(t)}{10~\rm M_{\odot}\rm yr^{-1}}\right),
    \label{two-component-model}
\end{equation}
where ${\rm Rate}(t)$ is the transient rate (e.g., SN Ia), $A$ and $B$ are dimensionless constants  that scale the relative contributions of the delayed and prompt components, respectively \citep{2005ApJ...629L..85S}. This empirical framework was initially established through observational constraints by \citep{2005A&A...433..807M} and theoretically formalized by \citep{2005ApJ...629L..85S}. It has since been extensively verified by large-scale surveys (e.g., \citealt{2006ApJ...648..868S,2008ApJ...682..262D}) and is now understood as a natural consequence of the power-law DTD of SNe Ia \citep{2012PASA...29..447M,2014ARA&A..52..107M}. In this framework, the transient rate is modeled as a linear superposition of a ``delayed" component (originating from old stellar populations and proportional to the total accumulated stellar mass, commonly seen in elliptical galaxies) and a ``prompt" component (arising from rapidly evolving progenitors in young populations and proportional to the current SFR, prevalent in star-forming galaxies). Ultimately, the event rates of SNe Ia and similar transients are jointly determined by both the stellar mass (representing the past integrated history) and the SFR (representing the present activity).

\subsection{Galaxies Sample} \label{subsec:sample selection}

To calculate the sky distributions of stellar mass and SFR of galaxies, a highly complete and extensive galaxy catalog is required. For this purpose, we adopted the Revised Galaxy List for the Advanced Detector Era (REGALADE; \citealt{2026A&Aregalade}), a comprehensive, all-sky galaxy compilation extending to 2000 Mpc. REGALADE integrates current major galaxy catalogs (see \citealt{2026A&Aregalade} for more details). After rigorously removing stellar contamination by using the Gaia catalog and visual inspection, the final REGALADE compilation contains nearly 80 million galaxies within 2000 Mpc, providing robust stellar mass estimates for 88\% of the sample \citep{2026A&Aregalade}. In addition to stellar masses, REGALADE provides Right Ascension (RA), Declination (DEC), redshifts, luminosity distances, and multi-band photometry, which are essential for deriving the SFRs of galaxies. However, it should be emphasized that despite its unprecedented size, REGALADE is inherently heterogeneous; some galaxies still lack measured complete multi-band coverage in the nearby universe (e.g., galaxies in the direction of Large Magellanic Cloud and Small Magellanic Cloud). On the other hand, because it merges various surveys with distinct scientific goals and observational strategies, the catalog exhibits non-uniform sky coverage and varying detection depths, particularly suffering from severe incompleteness near the Galactic plane and bulge region. Although REGALADE removed the contaminants very carefully, it was still not able to completely eliminate all the pollutants.

\begin{figure*}
 \includegraphics[width=1\linewidth]{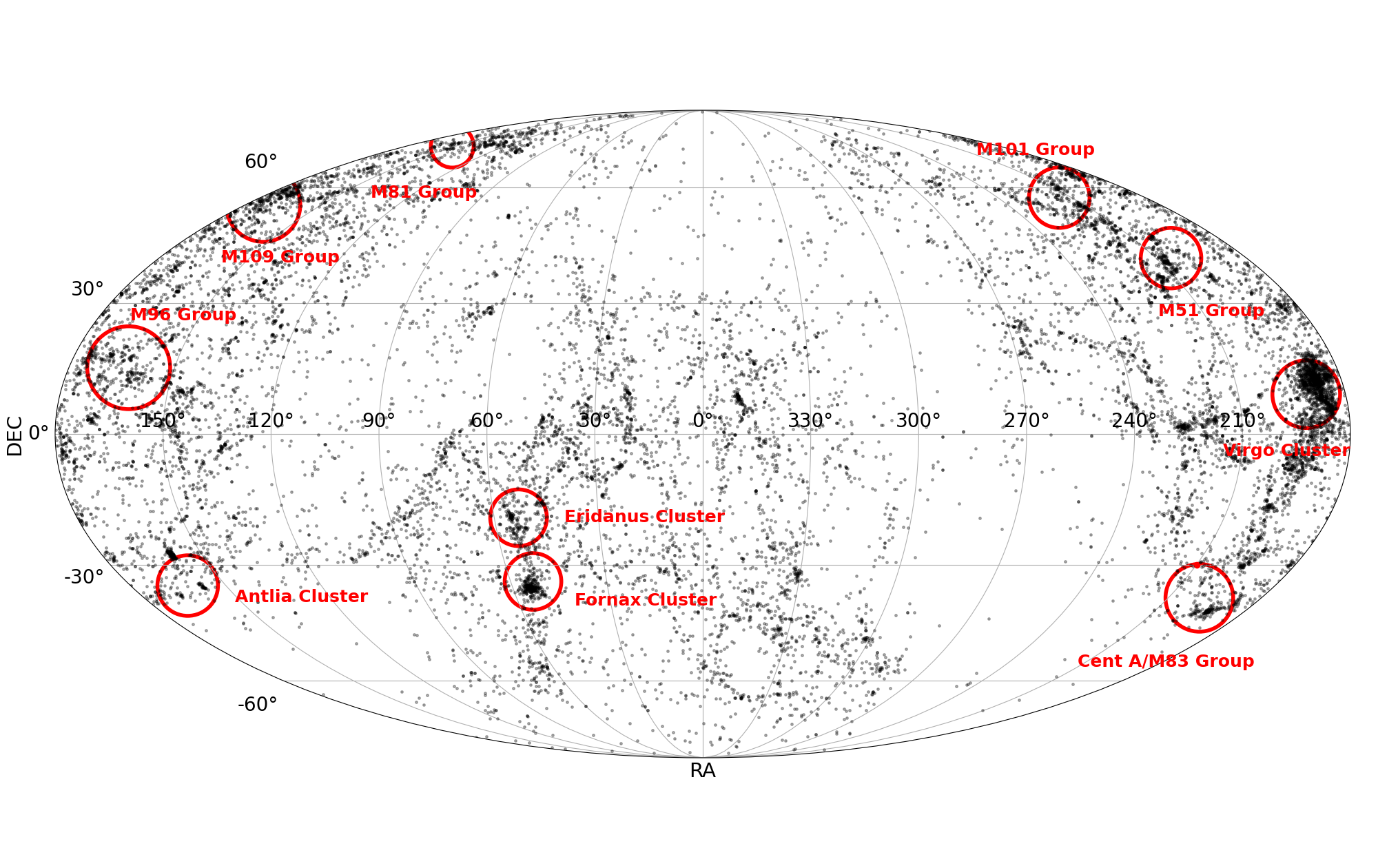}
 \caption{Sky distribution of galaxies presented in equatorial coordinates for a luminosity distance threshold of 50 Mpc, with red circles marking the approximate locations of prominent galaxy groups and clusters.}
 \label{galaxy cluster}
\end{figure*}

In Fig.~\ref{galaxy cluster}, we show the sky distribution of galaxies in equatorial coordinates within a luminosity distance threshold of 50 Mpc, with several galaxy groups and clusters marked by red circles. We can see that galaxy distributions in the local universe display pronounced clustering structures.

\subsection{Prediction of SFR of Galaxies}

\subsubsection{Sample Selection}\label{sample selection}
Since the REGALADE catalog provides stellar mass but does not include SFRs, we need to estimate the SFR for each galaxy through machine learning. \cite{2018ApJ...859...11S} provided the GALEX-SDSS-WISE Legacy Catalog 2 (GSWLC-2). Unlike earlier catalogs that rely solely on ultraviolet (UV) /optical photometry or empirical emission-line calibrations \citep{Salim_2016}, GSWLC-2 utilizes an advanced IR-luminosity-constrained, energy-balance SED fitting method. By explicitly incorporating the total infrared luminosity derived from Wide-field Infrared Survey Explorer (WISE, \citealt{2010AJ....140.1868W,allwise2019}), this approach strictly enforces energy balance between dust attenuation in the UV/optical and thermal emission in the infrared. This method effectively breaks the age-dust degeneracy and allows the dust attenuation curve (both slope and UV bump strength) to vary as free parameters, yielding highly robust and physically consistent estimates of stellar masses and SFRs across diverse galaxy populations \citep{2018ApJ...859...11S}. We extract data from GSWLC-2 based on three conditions that include \texttt{SED fitting flag = OK, flag\_uv $\neq$ 1 and flag\_midir $\neq$ 0}. These conditions guarantee that SFRs are derived from broad-band SED fitting covering both UV and mid-infrared wavelengths. Thus, we have approximately 380,000 galaxies with SFR extracted from GSWLC-2.

To construct a clean and reliable galaxy sample for our analysis, we applied several quality cuts to the REGALADE catalog \citep{2026A&Aregalade}. First, we required \texttt{ref\_D\_in < 12}, which selects galaxies whose best distance estimate originates from one of the first 11 input catalogs (ranked 0 to 11 in Table 1 of \citealt{2026A&Aregalade}). Second, we excluded sources with \texttt{f\_reliability = 1}. As defined in \cite{2026A&Aregalade}, this flag identifies galaxies that appear exclusively in one or two lower-reliability catalogs, which may be more prone to spurious detections or artifacts. Third, we restricted the sample to galaxies with stellar masses in the range \texttt{$5< {\rm log}M<12.5$}, excluding both low-mass systems with highly uncertain mass estimates and potentially spurious high-mass outliers. Fourth, we discarded galaxies with \texttt{griz\_ref = 5 and 7}. For these sources, the optical magnitudes were not retrieved from the standardized Legacy Surveys \citep{2022RAA....22f5001Z,2024ApJS..272...39W}, DELVE \citep{2022ApJS..261...38D}, or Pan-STARRS \citep{2016arXiv161205560C_ps1} used for the bulk of the REGALADE sample; thus, applying the ``mass-to-light" ratio to these objects may introduce systematic biases \citep{2026A&Aregalade}. After applying these filtering criteria, we obtained a parent sample of more than 33 million galaxies from REGALADE for further analysis.

To train and validate our machine-learning models that predict the SFRs of galaxies, we perform the cross-match on \texttt{TOPCAT} \citep{2005ASPC..347...29T} between REGALADE and GSWLC-2 with 2 arcsec error box to derive the SFRs of part of galaxies. We adopted a stricter 2-arcsec threshold. Our primary motivation for this conservative choice is that this cross-matched subset serves as the training and validation set for our machine-learning models. In machine-learning applications, the purity and reliability of the training labels are paramount. A 2-arcsec radius is stringent enough to effectively eliminate ambiguous matches and spurious associations (e.g., chance alignments with foreground stars or background sources), while still accommodating the typical astrometric uncertainties between optical and mid-infrared emission centroids. This strict criterion ensures that our models are trained on a highly reliable, less-contaminated dataset. As a result, in the cross-matching process, we obtained approximately 335,000 valid galaxies that contain SFR. These galaxies can be used as the training set and validation set for the machine learning model.

\subsubsection{Training and validation of machine learning model}
The use of machine learning to predict galaxy parameters and properties has been widely explored in literature. For example, \cite{2019A&A...622A.137B} employed a Random Forest model on data from the Sloan Digital Sky Survey (SDSS) and the WISE to estimate SFRs and stellar masses. Similarly, \cite{2019MNRAS.486.1377D} utilized the same algorithm to predict specific SFRs using SDSS data alone. Expanding into deep learning, \cite{2020MNRAS.493.4808S} successfully predicted SFRs and stellar masses from the Galaxy And Mass Assembly (GAMA) survey. More recently, \cite{Z24} (hereafter Z24) applied the CatBoost model \citep{2017arXiv170609516P} to combined SDSS and WISE datasets for identical parameter predictions.

Taking into account the limitations of feature combinations and the accuracy of predictions, we follow the methodology of Z24 and utilize the CatBoost model to predict the SFRs. To evaluate the performance of our predictions, we adopt several statistical metrics following Z24. The root mean squared error (RMSE) is defined as

\begin{equation}
    {\rm RMSE}= \sqrt{\frac{1}{m}\sum_{j=0}^{m-1} \left(S_{p,j}-S_{t,j} \right)^2 } ,
\end{equation}
where $m$ denotes the total number of galaxies (sample size), while $S_{p,j}$ and $S_{t,j}$ represent the predicted and true values of $\rm logSFR$ for the $j$-th galaxy, respectively. The standard deviation of the residuals is calculated via

\begin{equation}
    \sigma = \sqrt{ \frac{1}{m}\sum _{j=0}^{m-1}\left( \Delta S_j - \overline{\Delta S_j} \right)^2} .
\end{equation}
Additionally, the systematic offset between the predicted and actual values is defined as the bias, expressed as ${\rm bias} = \langle \Delta S \rangle $, where the residual is $\Delta S =S_p - S_t$. Predictions satisfying $|\Delta S|>3\sigma$ are classified as outliers, and the outlier fraction, $\eta$, is defined as the percentage of these catastrophic outliers \citep{Z24}.

\begin{figure*}
    \centering 
    \subfigure[Training result]{
        \includegraphics[width=3.4in]{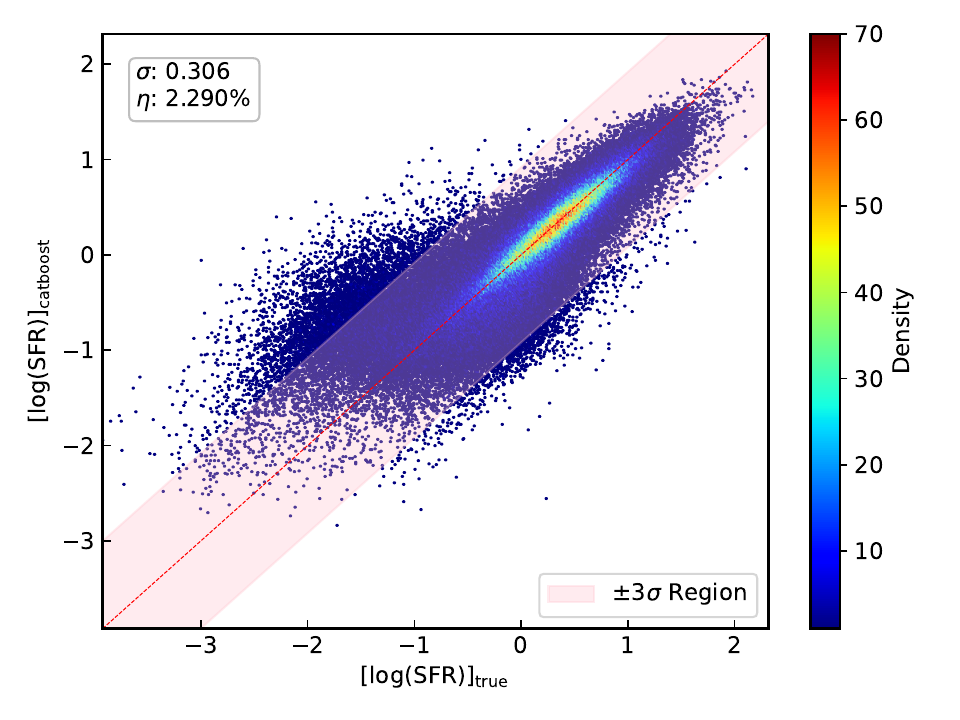}
    }
    \subfigure[Training result residual]{
        \includegraphics[width=3.4in]{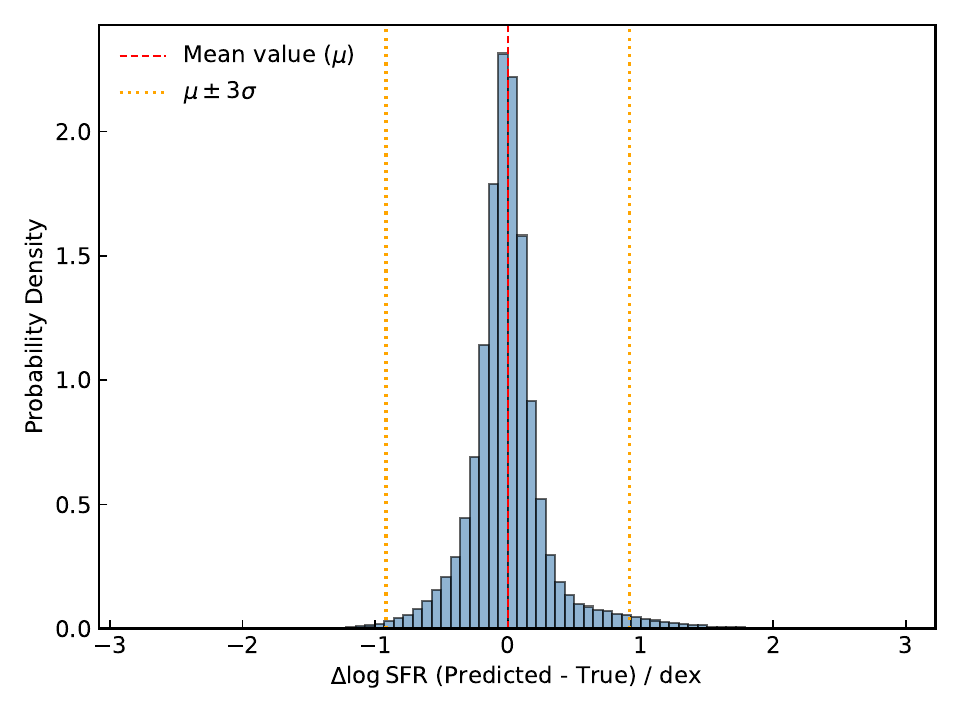}
    }  
    \\
    \subfigure[Blind test]{
        \includegraphics[width=3.4in]{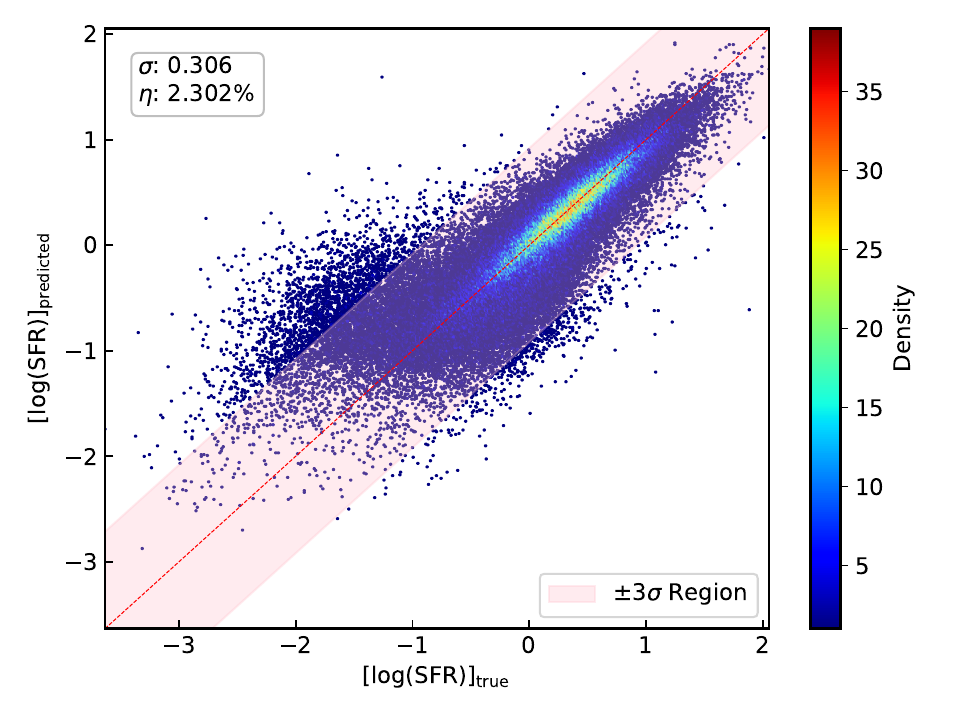}
    }
    \subfigure[Blind test residual]{
        \includegraphics[width=3.4in]{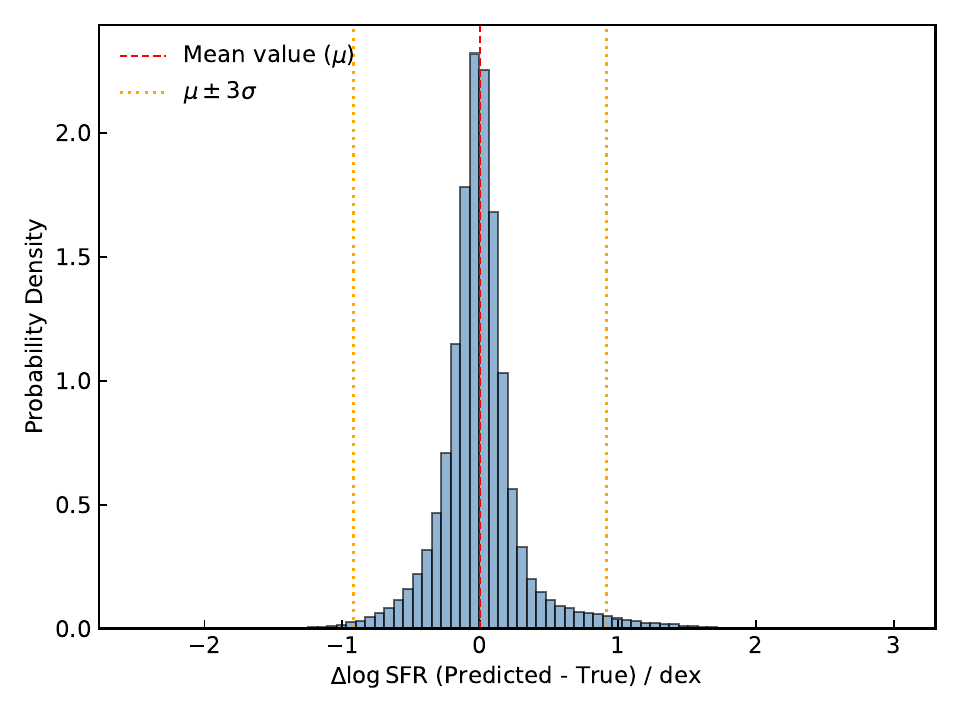}
    }
    \caption{Comparison of predicted $\rm logSFR$ and corresponding residuals for the training (top row) and blind test (bottom row) datasets. Left panels: Predicted versus true $\rm logSFR$ values, where the red dashed line denotes the 1:1 relation and the pink shaded area represents the $\pm3\sigma$ region. Right panels: Probability density distributions of the residuals. The vertical red dashed line indicates the mean residual ($\mu$), and the yellow dotted lines mark the $\mu \pm 3\sigma$ boundaries.}
    \label{fig:training and blind test}
\end{figure*}

We construct a feature combination that included $g$, $g-r$, $r-z$, $z-W1$, $W1-W2$, redshift, and stellar mass ($\log M_\ast$). Utilizing the CatBoost model, we applied a 10-fold cross-validation approach to model training and validation. Two-thirds of the sample were allocated for training and validation, while the remaining one-third was reserved as a blind test set to verify the predictive accuracy of the model. The training and blind test results are illustrated in Fig.~\ref{fig:training and blind test}. Panels (a) and (c) present the comparisons between the true and predicted values for the training and blind test sets, respectively, while Panels (b) and (d) show their corresponding residual probability density distributions. A detailed performance comparison with Z24 is summarized in Table~\ref{table:metrics_comparison}. Remarkably, despite incorporating fewer photometric magnitudes than Z24, our model achieves a significantly lower RMSE of 0.306 compared to their 0.349, largely owing to the inclusion of redshift and stellar mass in our feature combination. Furthermore, our prediction residuals closely follow a Gaussian distribution. Although our outlier fraction ($\eta \approx 2.3\%$) is approximately 1.2 percentage points higher than that of Z24 ($\eta = 1.1\%$), this discrepancy is acceptable given that our sky distributions represent a cumulative projection along the line of sight.

\subsubsection{Predicting the SFR of galaxies}
Since the universe is homogeneous on a large scale, we focus on the distribution of galaxies in the nearby universe. Therefore, in addition to the filtering conditions in Section~\ref{sample selection} , we set a luminosity distance threshold of 200 Mpc to calculate the distribution of stellar mass and SFR of galaxies. Our initial catalog comprises 364,148 galaxies within 200 Mpc. By cross-matching this sample with the GSWLC-2 catalog using a 2-arcsec matching radius, we successfully obtained direct SFR measurements for 38,341 galaxies, thereby bypassing the need for further machine learning prediction. Consequently, the remaining 325,807 galaxies, which lack pre-existing measurements, are designated as the target sample for our machine learning SFR predictions.
Utilizing the trained CatBoost model described above, we successfully predicted the SFRs for the remaining 325,807 galaxies. Combined with the cross-matched data, we have now established a complete census of SFRs for all 364,148 galaxies within 200 Mpc. This comprehensive dataset enables us to investigate the sky distributions of these galaxies in the subsequent analysis.

\subsection{Calculation of the Sky Distributions of Galaxies}

Next, we visualize the sky distributions of stellar mass and SFR of galaxies using the Python package \texttt{healpy} \citep{healpy}, which is the python version of \texttt{HEALPix} \citep{healpix}. 
In \texttt{HEALPix}, the celestial sphere can be divided into curvilinear quadrilaterals of identical area for all grids based on a given angular resolution of the grid, and each grid has a fixed number and center coordinates \citep{healpix}.

The selection of the angular resolution of the grid depends on the FoV of the telescope. We use \texttt{ang2pix} to convert the coordinates of galaxies to the fixed grid number and identify the location of each galaxy on the celestial sphere. Then we can calculate the total stellar mass / SFR within each grid for a given maximum distance, 
and finally obtain the full sky distributions of stellar mass / SFR of galaxies.
 
\setlength{\tabcolsep}{12pt}
\begin{deluxetable*}{lccccc}
  \tablecaption{Model performance comparison\label{table:metrics_comparison}}
  \tablecolumns{6}
  
  \startdata \\
    \textbf{Data} & \textbf{RMSE} & \textbf{$\sigma$} & \textbf{bias} & \textbf{$\eta$} & \textbf{Reference}  \\
    \hline
    REGALADE & 0.306 & 0.306 & -0.0001 & 2.290~\% & Model Training  \\
    REGALADE & 0.306 & 0.306 & -0.0015 & 2.302~\% & Blind Test \\
    SDSS+WISE & 0.349 & 0.35 & 0.002 & 1.1~\% &  Z24 \\
\enddata
\tablecomments{Evaluation metrics comparison between model training, blind test and \citealt{Z24}.}
\end{deluxetable*}
\setlength{\tabcolsep}{6pt} 

\section{Results} \label{sec:results}

In this section, we present the results of the sky distributions of stellar mass and SFR of galaxies in the local universe. 
Our method enables the calculation of sky distributions at various angular resolutions, determined by the telescope’s field of view (FoV) and the distance threshold. 
Here, we used the FoVs of Multi-channel Photometric Survey Telescope (Mephisto\footnote{{\href{http://www.mephisto.ynu.edu.cn/}{http://www.mephisto.ynu.edu.cn/}}}, \citealt{2019gage.confE..14L,2024ApJ...969..126Y,2025ApJ...979...38C}), ZTF \citep{2019PASP..131a8002B} and the Einstein Probe  (EP, \citealt{2022arXiv220909763Y}) as examples to calculate the sky distributions of stellar mass and SFR of galaxies. 
The FoVs of Mephisto, ZTF and EP are 3.14 $\rm deg^2$, 47 $\rm deg^2$ and 3600 $\rm deg^2$, respectively. Based on the division of the celestial sphere mentioned above, we matched the angular resolution of the grid approximately with the square root of the FoV and selected angular resolutions of $1.83^{\circ}$, $7.33^{\circ}$ and $58.6^{\circ}$, corresponding to Mephisto, ZTF and EP, respectively. These resolutions were chosen to ensure that the area of the grid closely matches the FoV of the telescopes. 
In Table~\ref{theta_pix}, we present the angular resolutions of the grids used in our computations.
With the chosen angular resolutions of the grid, we can visualize the corresponding sky distributions of stellar mass and SFR of galaxies.

\begin{figure*}
\centering
\subfigure{
\begin{minipage}[b]{.3\linewidth}
\centering
\includegraphics[scale=0.16]{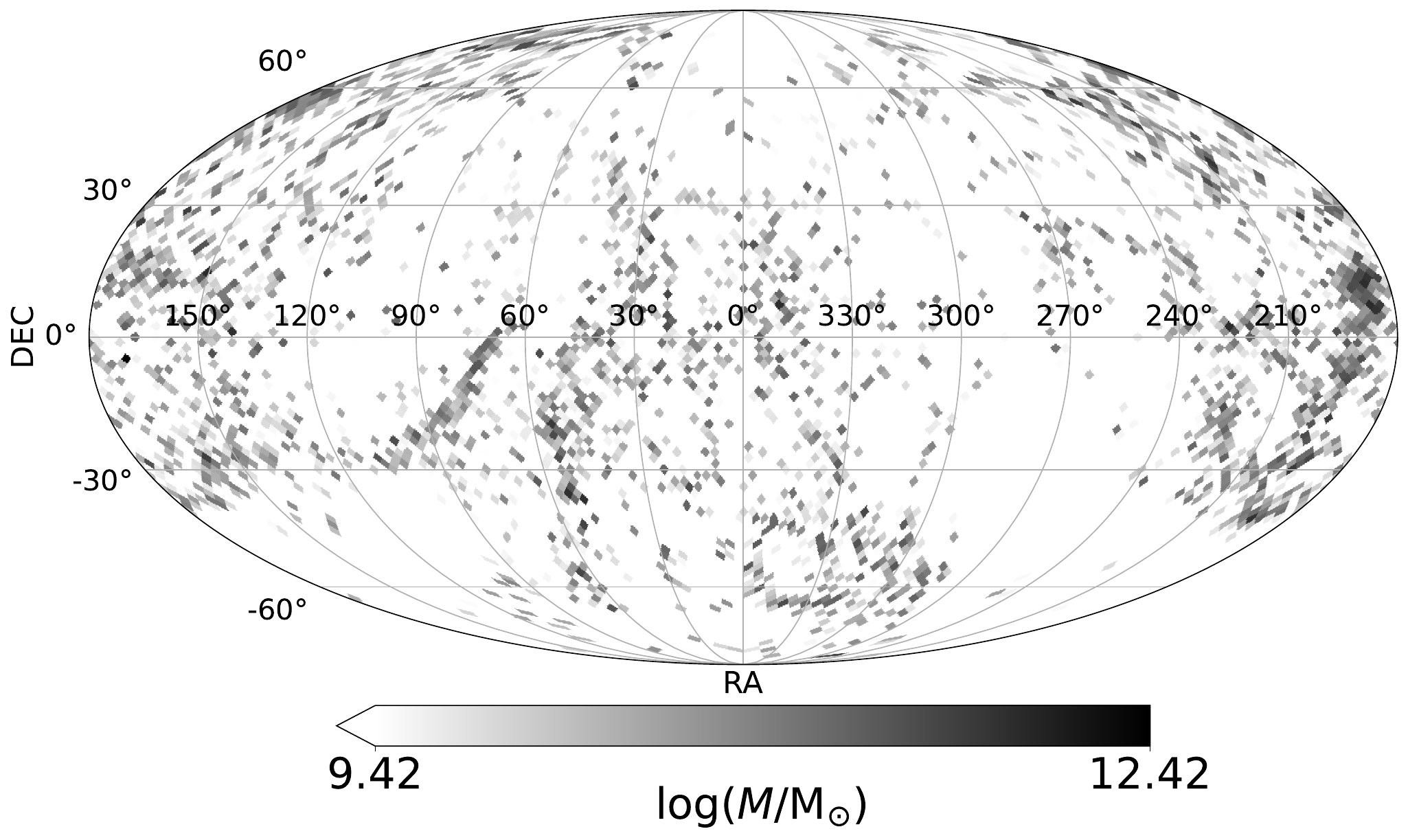}
\end{minipage}
}
\subfigure{
\begin{minipage}[b]{.3\linewidth}
\centering
\includegraphics[scale=0.16]{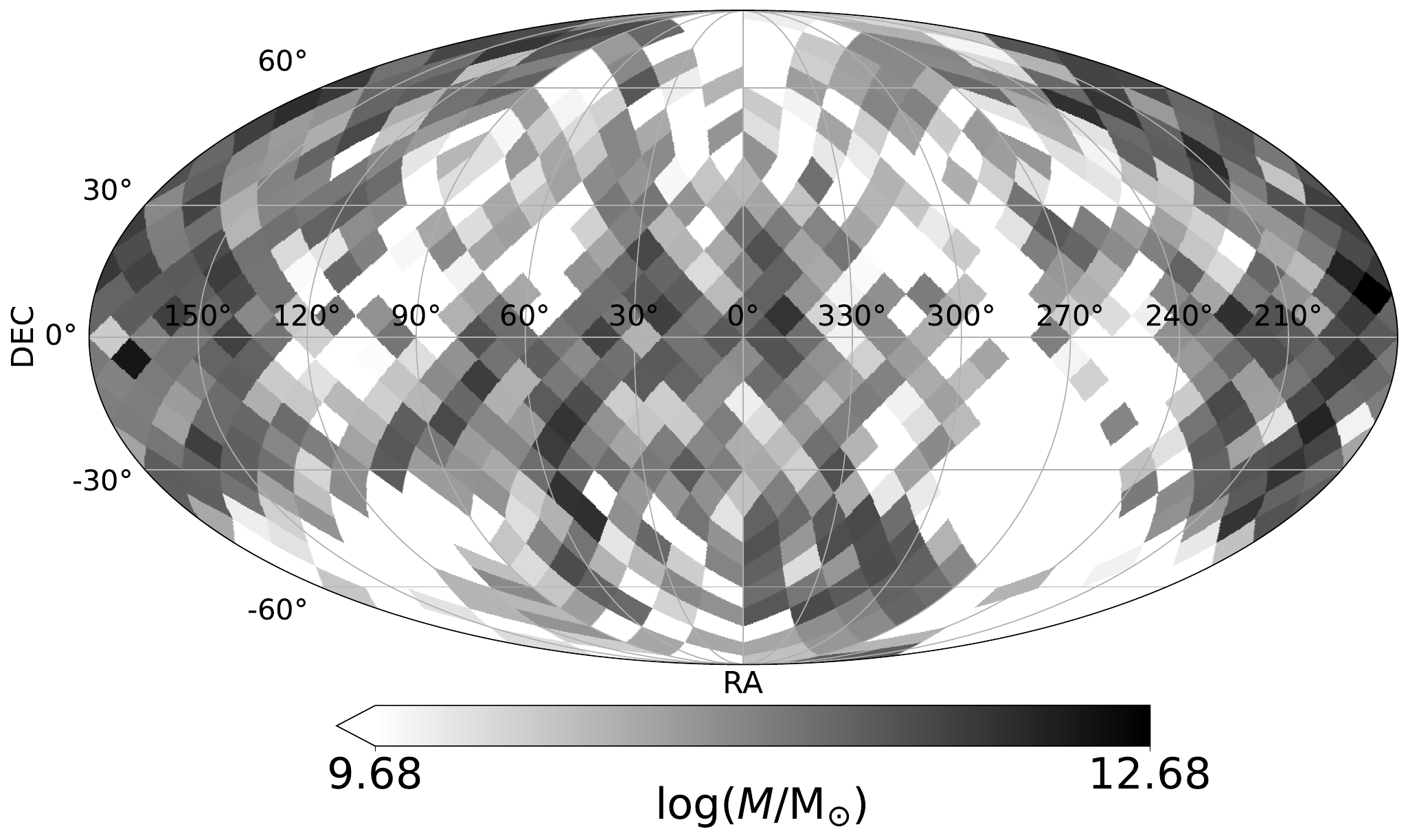}
\end{minipage}
}
\subfigure{
\begin{minipage}[b]{.3\linewidth}
\centering
\includegraphics[scale=0.16]{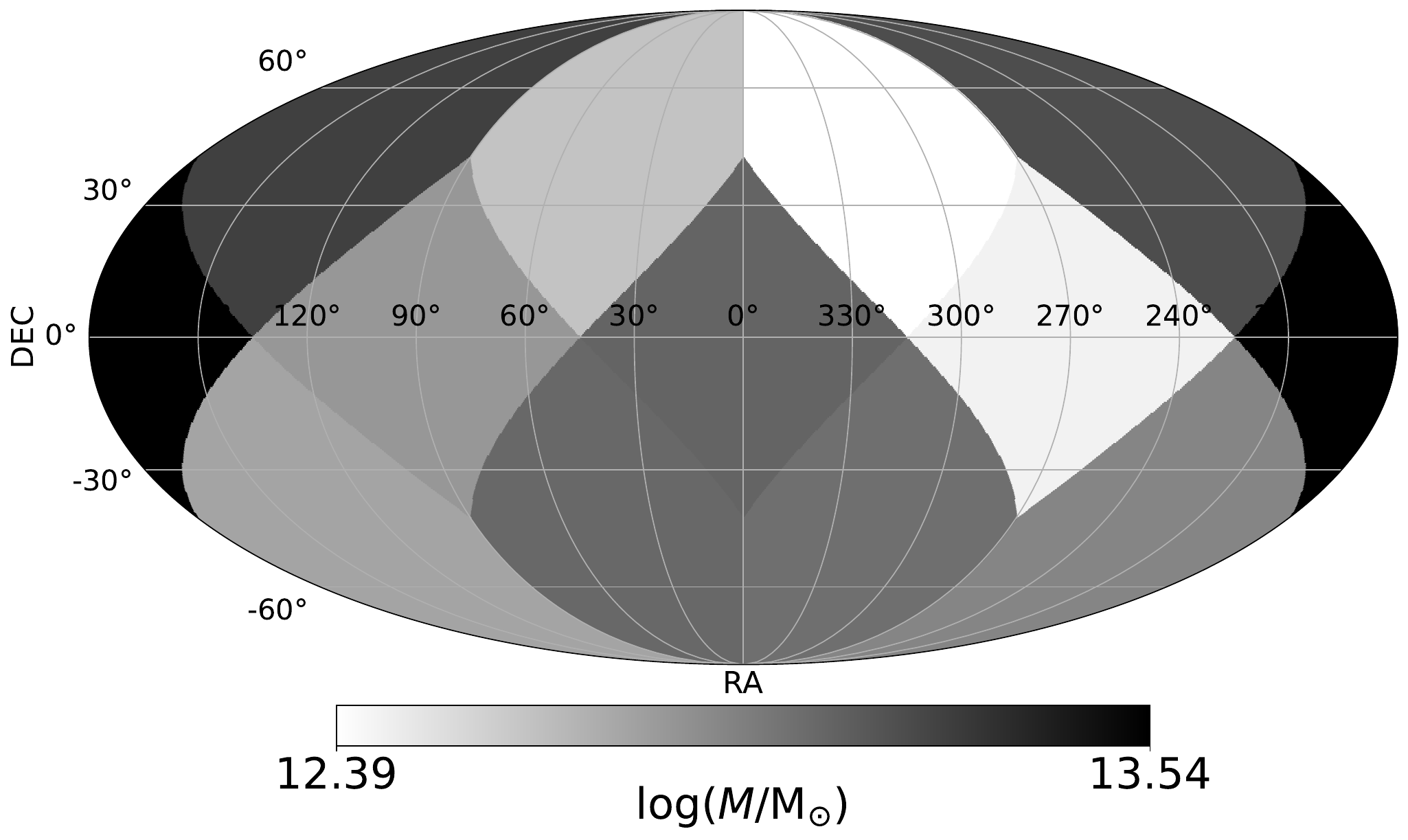}
\end{minipage}
}
\subfigure{
\begin{minipage}[b]{.3\linewidth}
\centering
\includegraphics[scale=0.16]{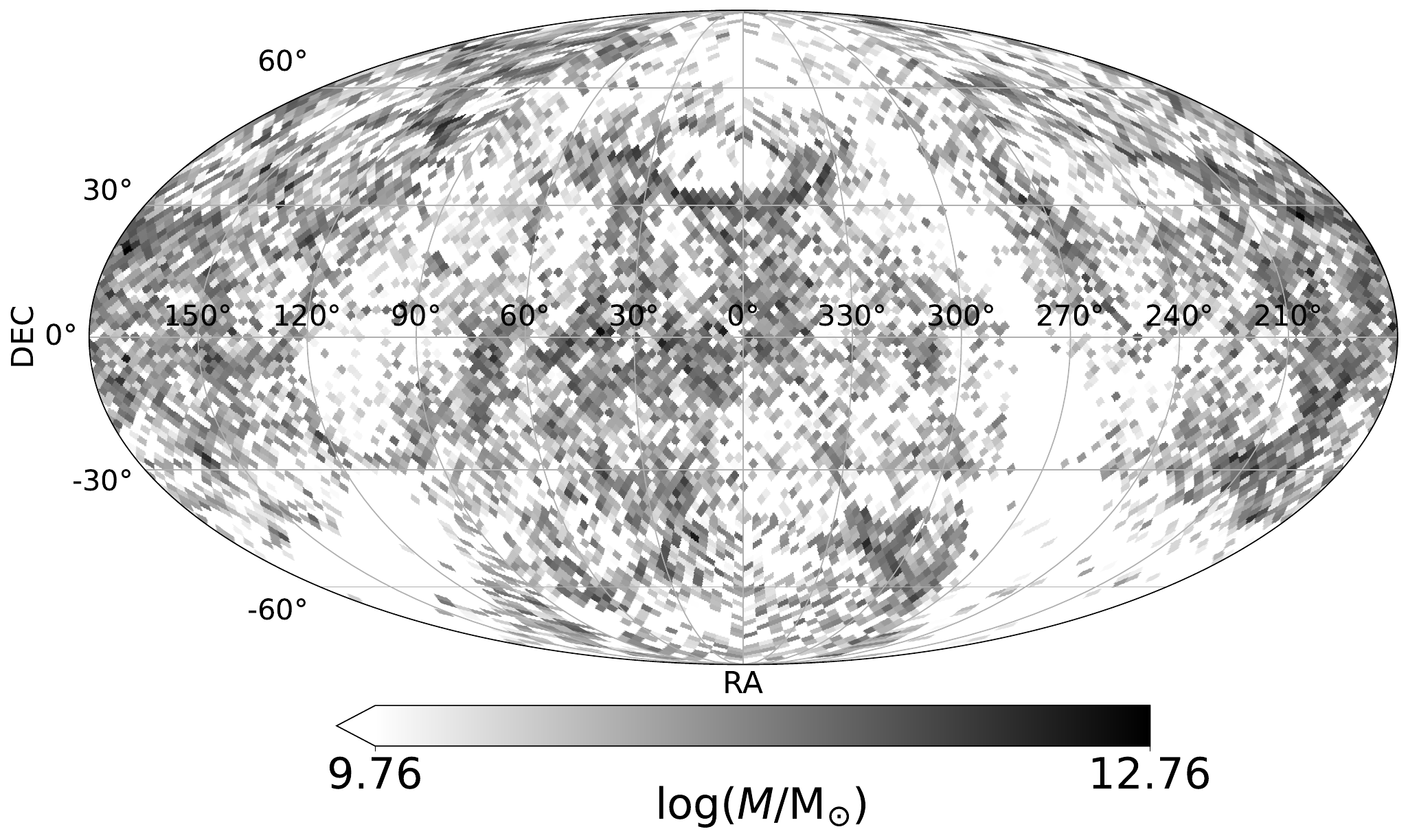}
\end{minipage}
}
\subfigure{
\begin{minipage}[b]{.3\linewidth}
\centering
\includegraphics[scale=0.16]{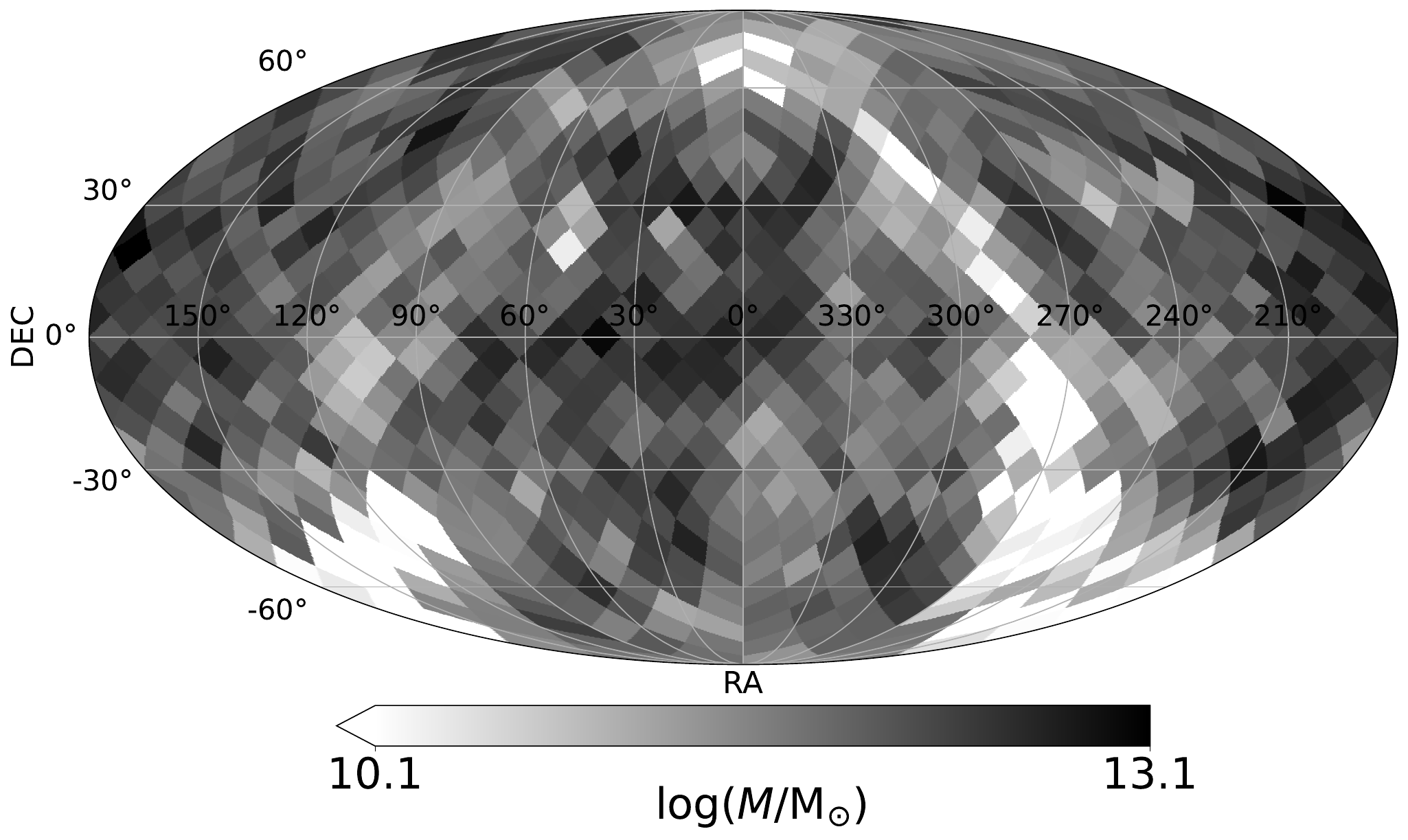}
\end{minipage}
}
\subfigure{
\begin{minipage}[b]{.3\linewidth}
\centering
\includegraphics[scale=0.16]{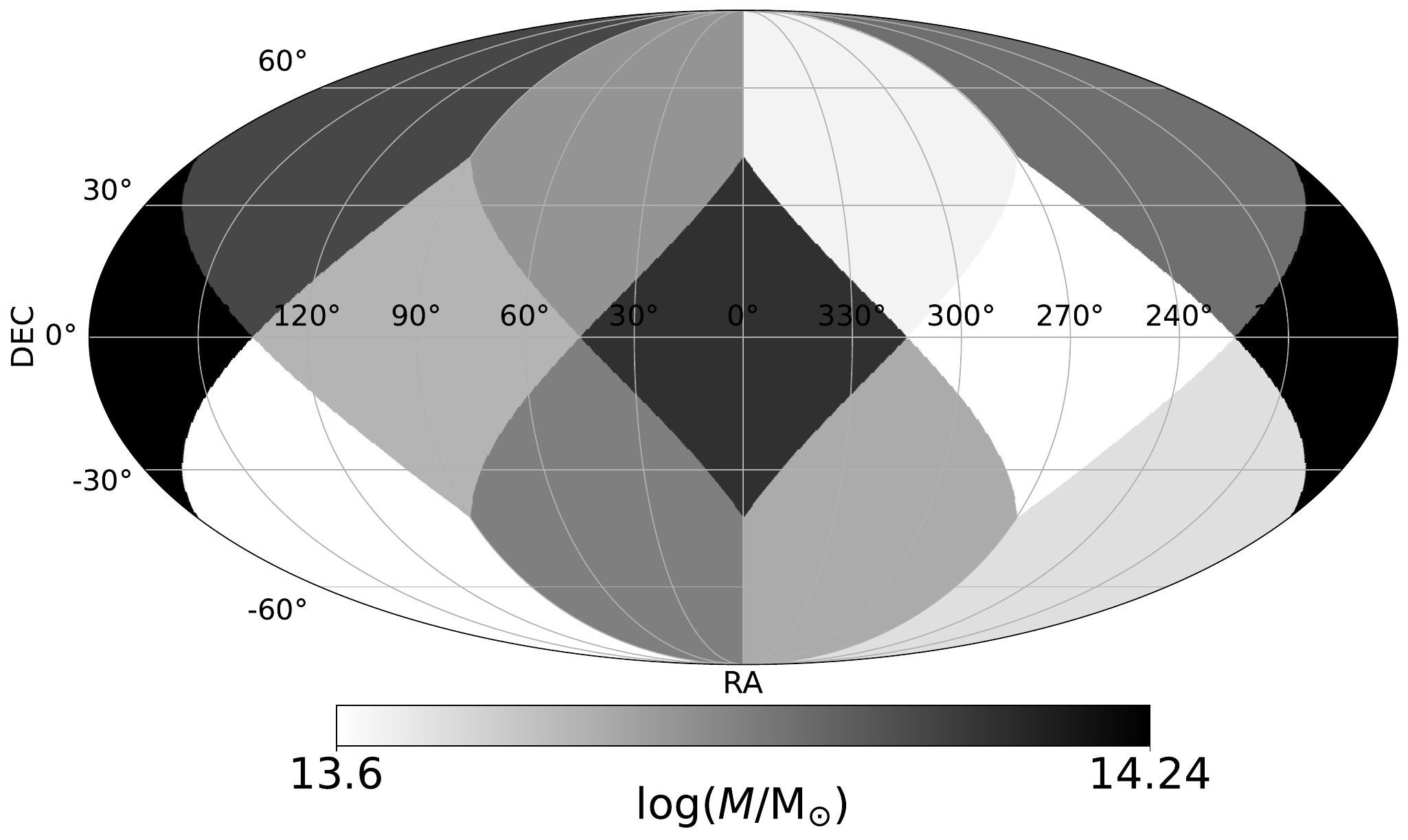}
\end{minipage}
}
\subfigure{
\begin{minipage}[b]{.3\linewidth}
\centering
\includegraphics[scale=0.16]{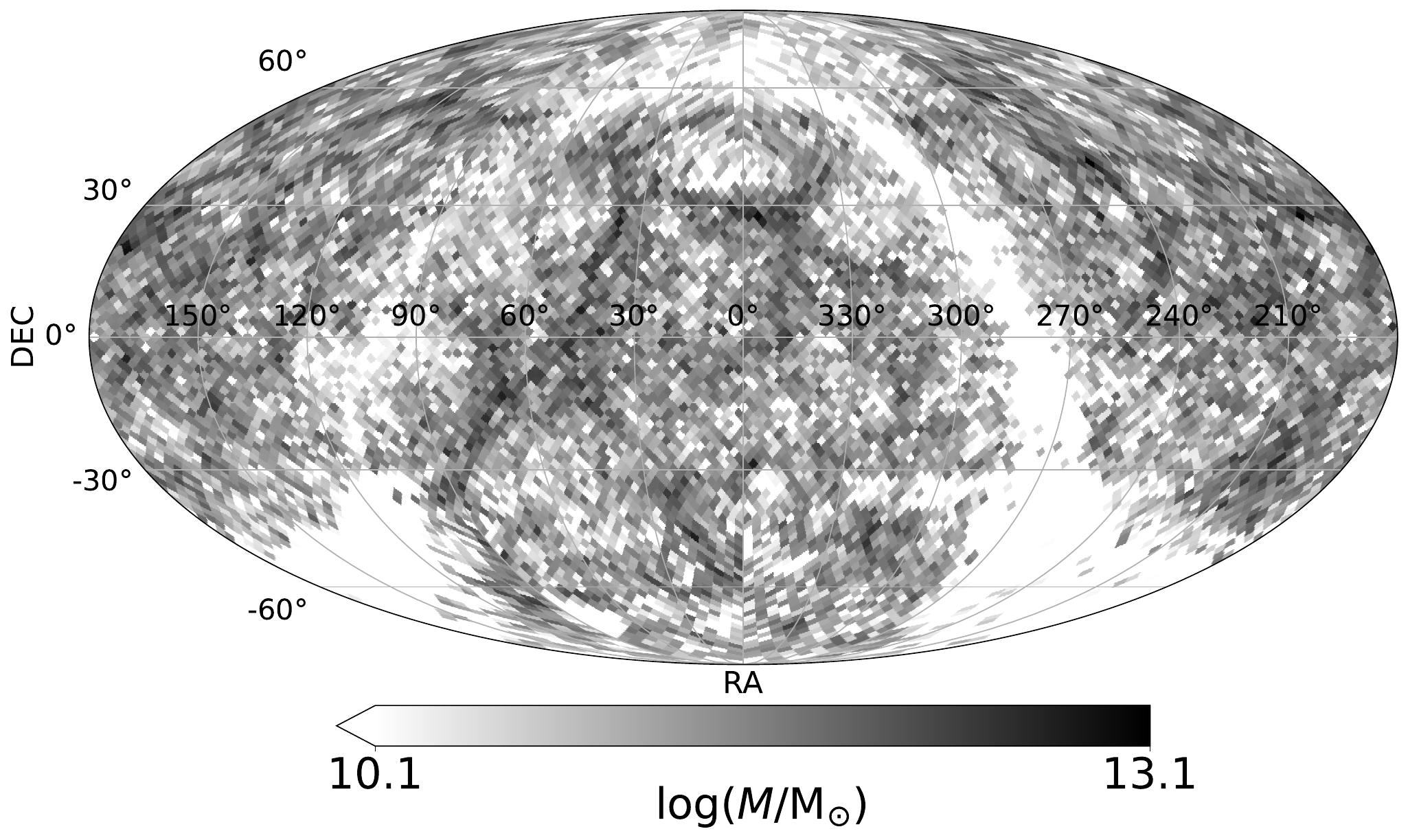}
\end{minipage}
}
\subfigure{
\begin{minipage}[b]{.3\linewidth}
\centering
\includegraphics[scale=0.16]{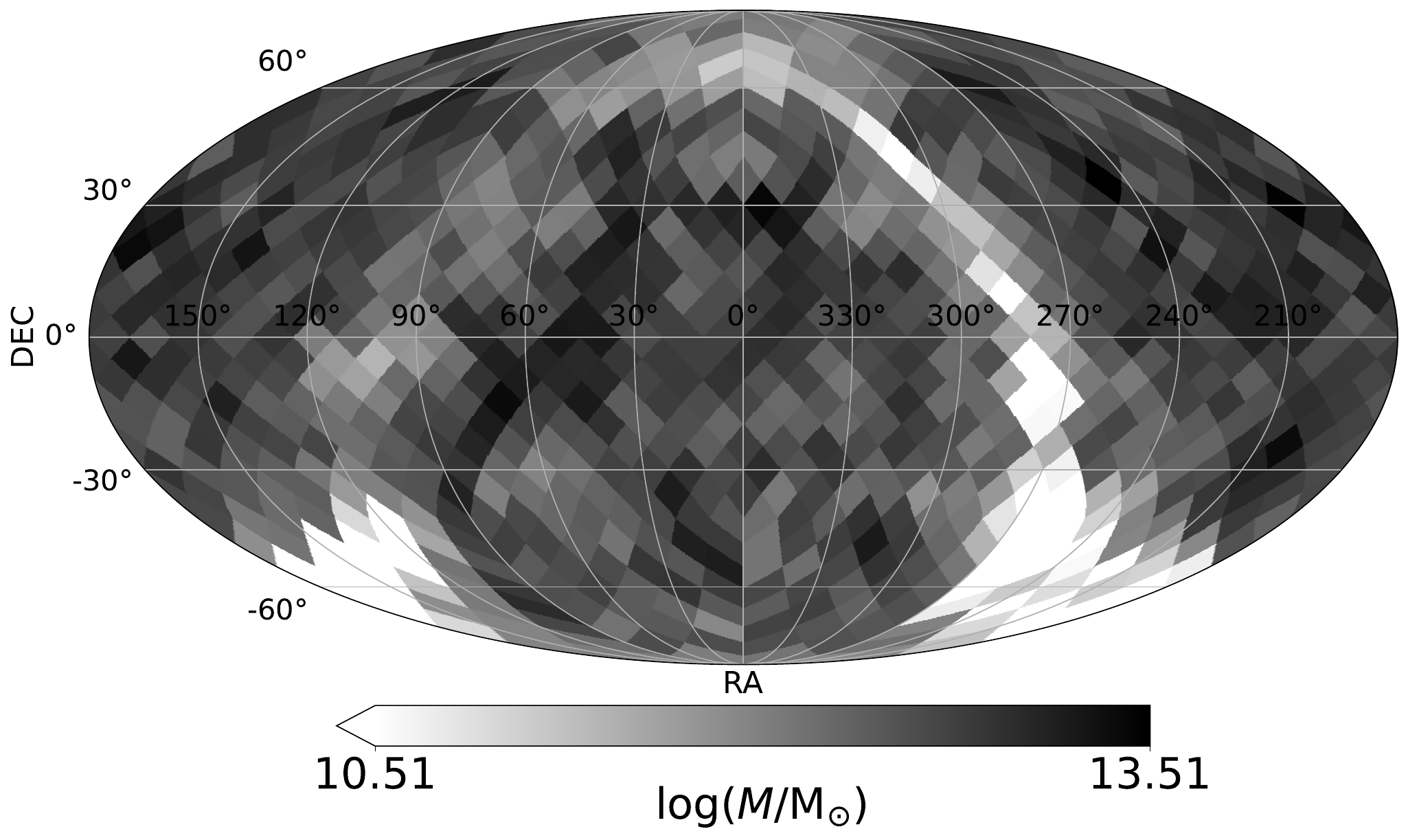}
\end{minipage}
}
\subfigure{
\begin{minipage}[b]{.3\linewidth}
\centering
\includegraphics[scale=0.16]{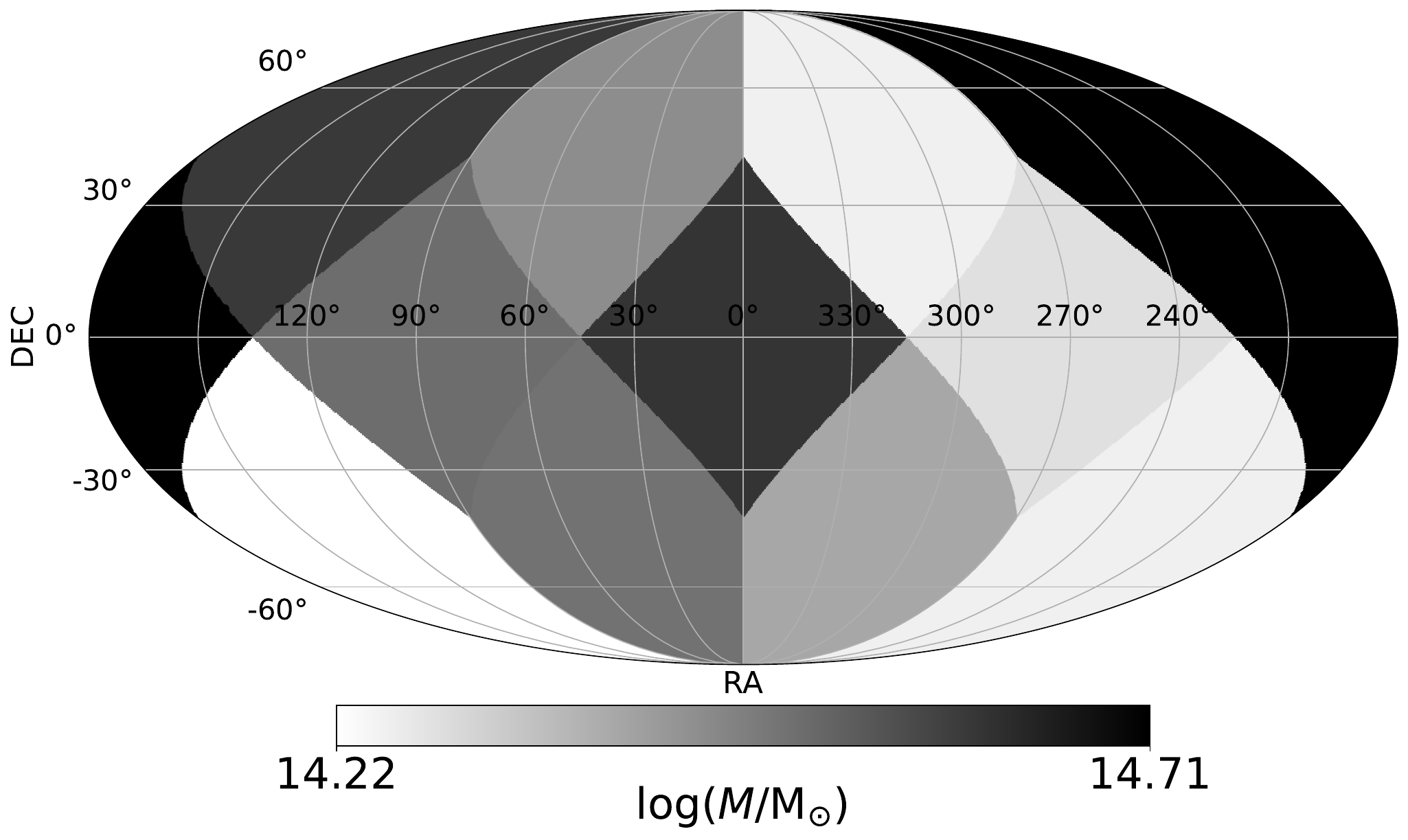}
\end{minipage}
}
\subfigure{
\begin{minipage}[b]{.3\linewidth}
\centering
\includegraphics[scale=0.16]{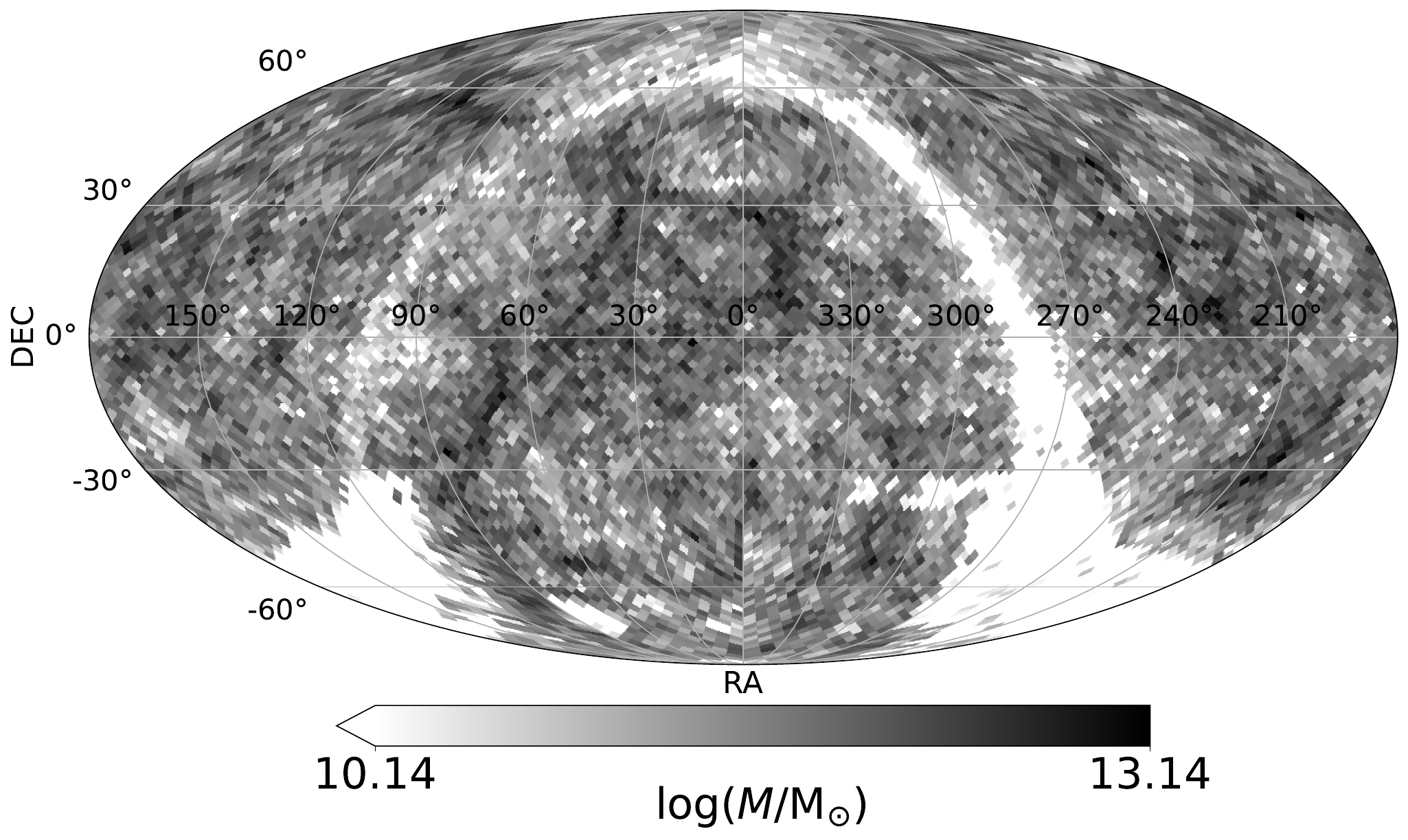}
\text{(a) $\theta=1.83^{\circ}$}
\end{minipage}
}
\subfigure{
\begin{minipage}[b]{.3\linewidth}
\centering
\includegraphics[scale=0.16]{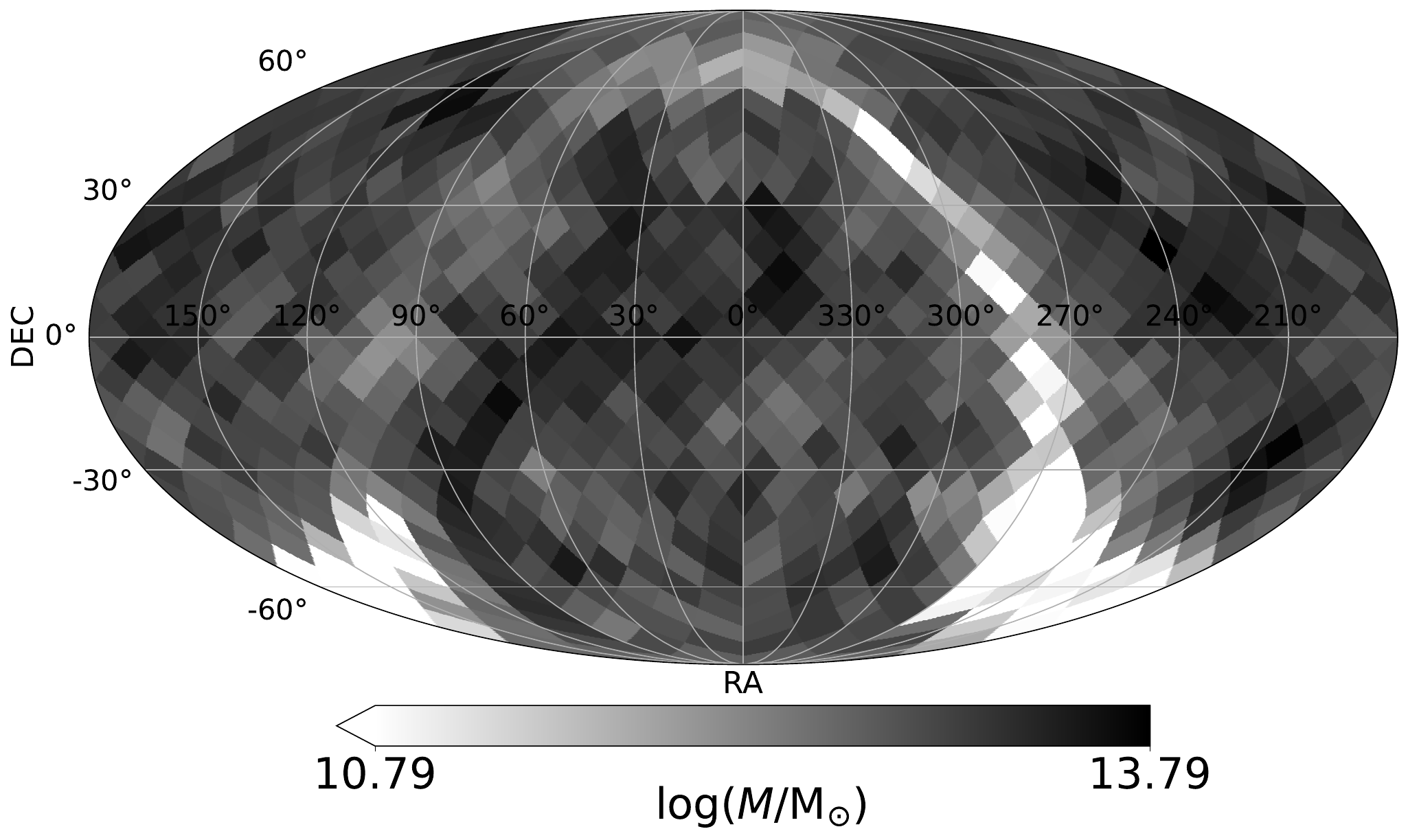}
\text{(b) $\theta=7.33^{\circ}$}
\end{minipage}
}
\subfigure{
\begin{minipage}[b]{.3\linewidth}
\centering
\includegraphics[scale=0.16]{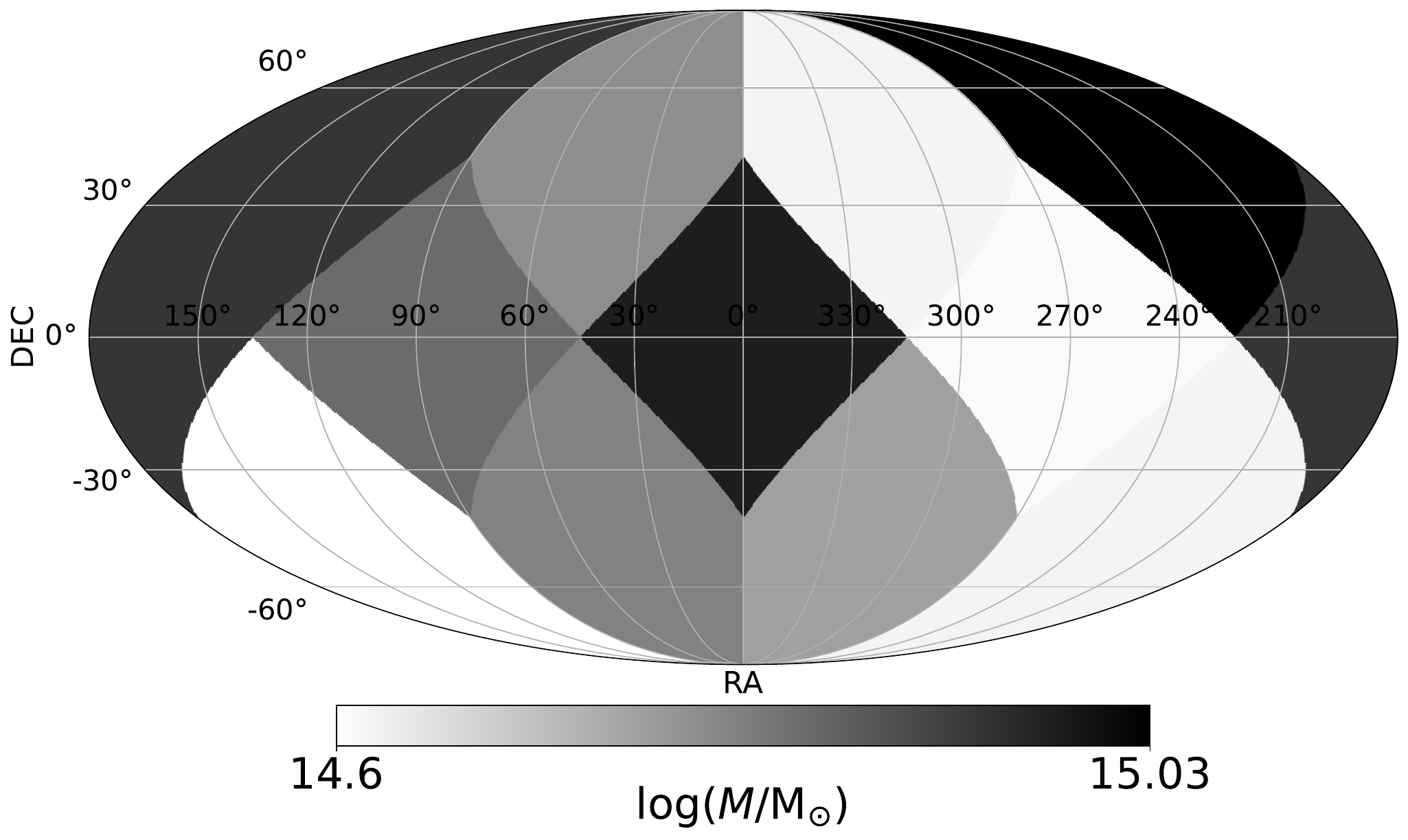}
\text{(c) $\theta=58.6^{\circ}$}
\end{minipage}
}
\caption{Sky distribution of stellar mass in equatorial coordinates. From left to right, the angular resolutions of grid of sky distributions at each column are $\theta=1.83^{\circ}$, $7.33^{\circ}$ and $58.6^{\circ}$, respectively. From top to bottom, the luminosity distance thresholds of sky distributions at each row are 50 Mpc, 100 Mpc, 150 Mpc, and 200 Mpc, respectively. The color bar range spans three orders of magnitude if the stellar mass range more than three orders of magnitude.}
\label{mass_distribution}
\end{figure*}

The sky distributions of stellar mass of galaxies are shown in Fig.~\ref{mass_distribution}, where columns (a), (b) and (c) correspond to angular resolution of the grid of $1.83^{\circ}$, $7.33^{\circ}$ and $58.6^{\circ}$, respectively. From top to bottom, the luminosity distance thresholds of the sky distributions of stellar mass in each row are 50 Mpc, 100 Mpc, 150 Mpc, 200 Mpc, respectively. 
In these sky distributions, darker colors indicate regions where the total stellar mass of galaxies within each grid cell is higher. 
We set the range of the color bar to span three orders of magnitude, meaning that the event rate and the probability of discovering transients in the darkest regions are about one thousand times higher than in the lightest regions.
The anisotropic structures are clearly visible in the sky distributions, highlighting the non-uniformity of galaxy populations at a specific luminosity distance threshold.
As the luminosity distance threshold
increases, the anisotropic distribution structures tend towards isotropy. 
In a survey project for the local universe, it is important to prioritize observations of the darker regions when aiming to efficiently discover transients associated with stellar mass in the nearby universe. Focusing on these regions increases the likelihood of detecting transients compared to targeting the lighter areas.

\begin{figure*}
\centering
\subfigure{
\begin{minipage}[b]{.3\linewidth}
\centering
\includegraphics[scale=0.16]{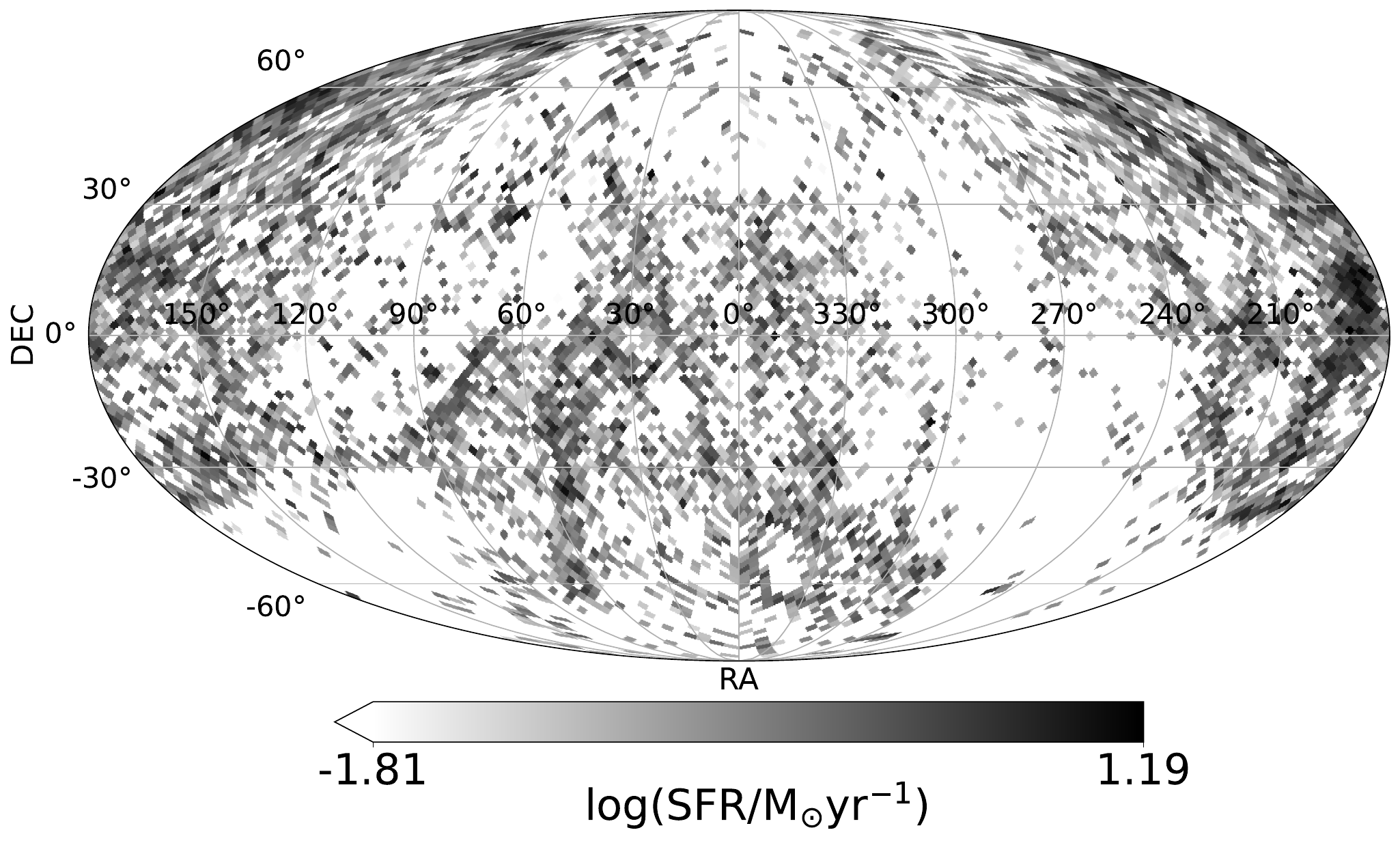}
\end{minipage}
}
\subfigure{
\begin{minipage}[b]{.3\linewidth}
\centering
\includegraphics[scale=0.16]{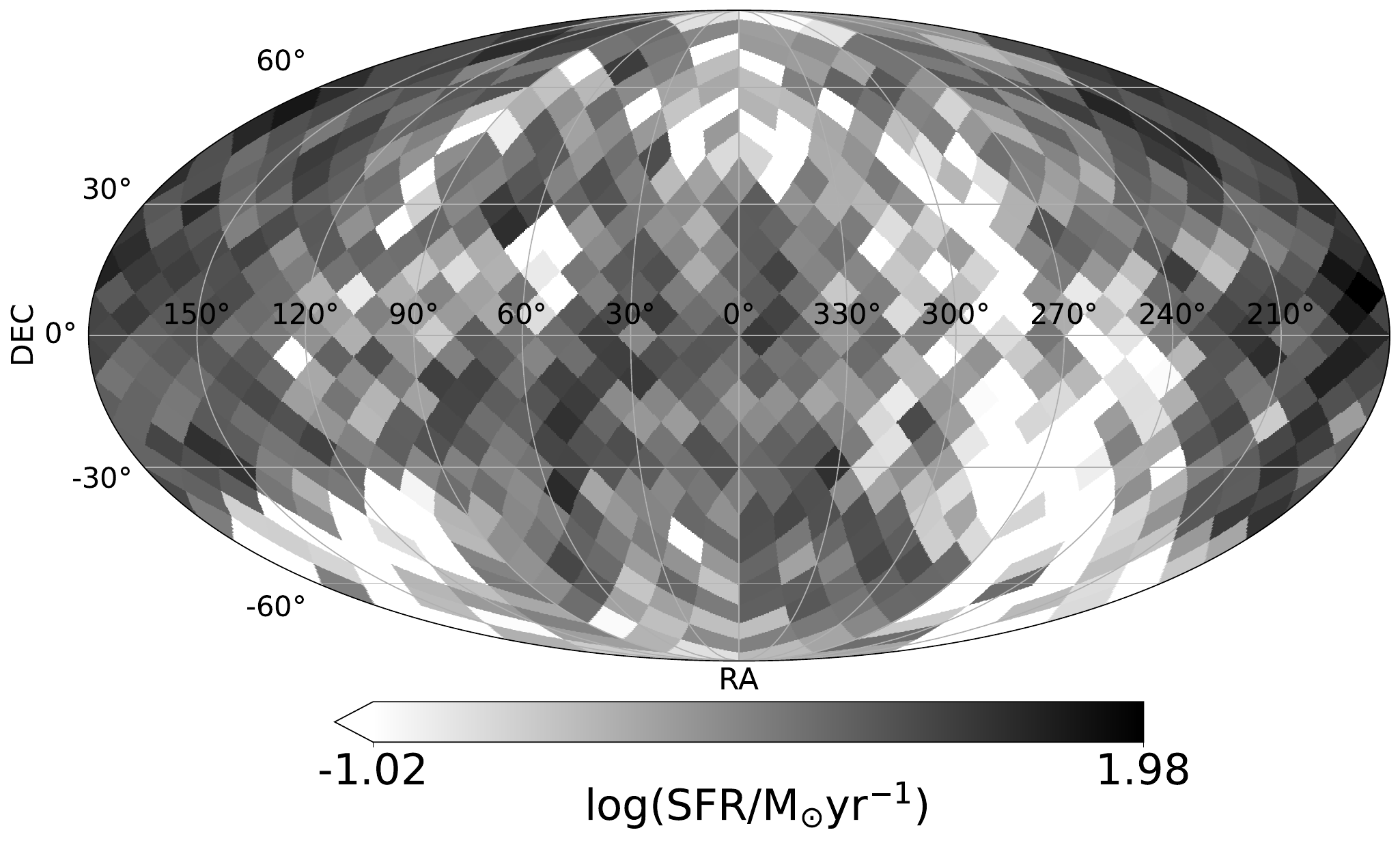}
\end{minipage}
}
\subfigure{
\begin{minipage}[b]{.3\linewidth}
\centering
\includegraphics[scale=0.16]{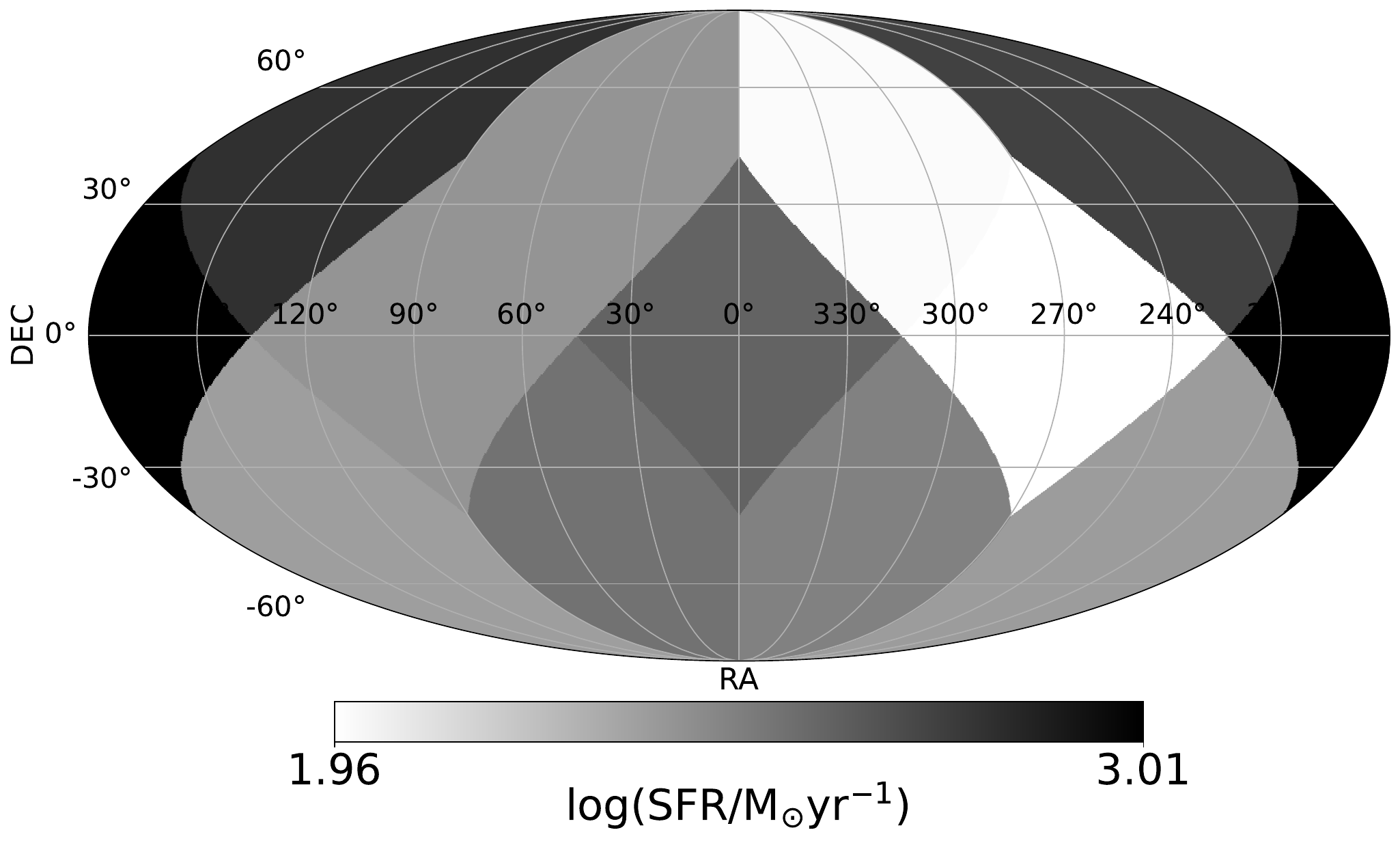}
\end{minipage}
}
\subfigure{
\begin{minipage}[b]{.3\linewidth}
\centering
\includegraphics[scale=0.16]{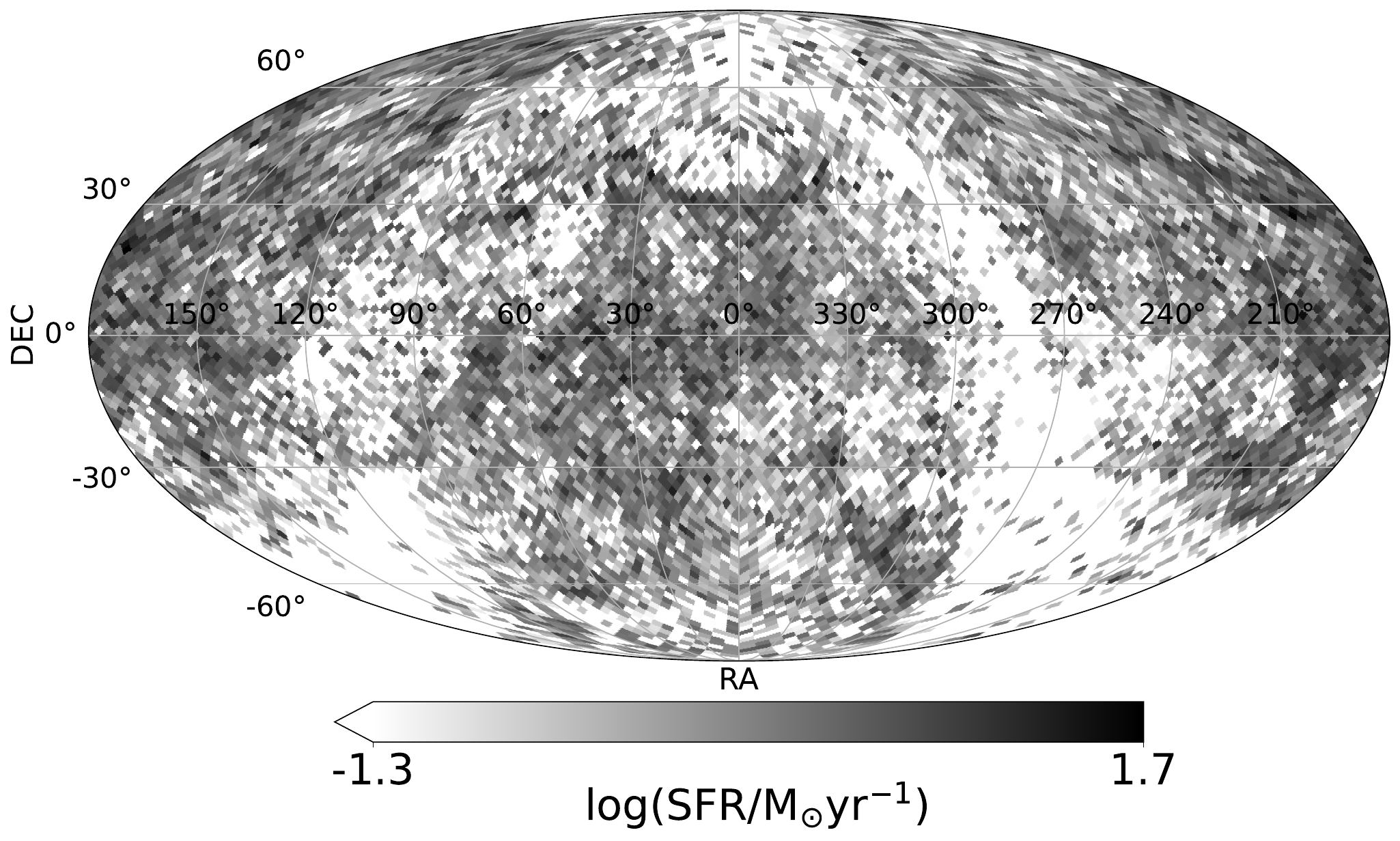}
\end{minipage}
}
\subfigure{
\begin{minipage}[b]{.3\linewidth}
\centering
\includegraphics[scale=0.16]{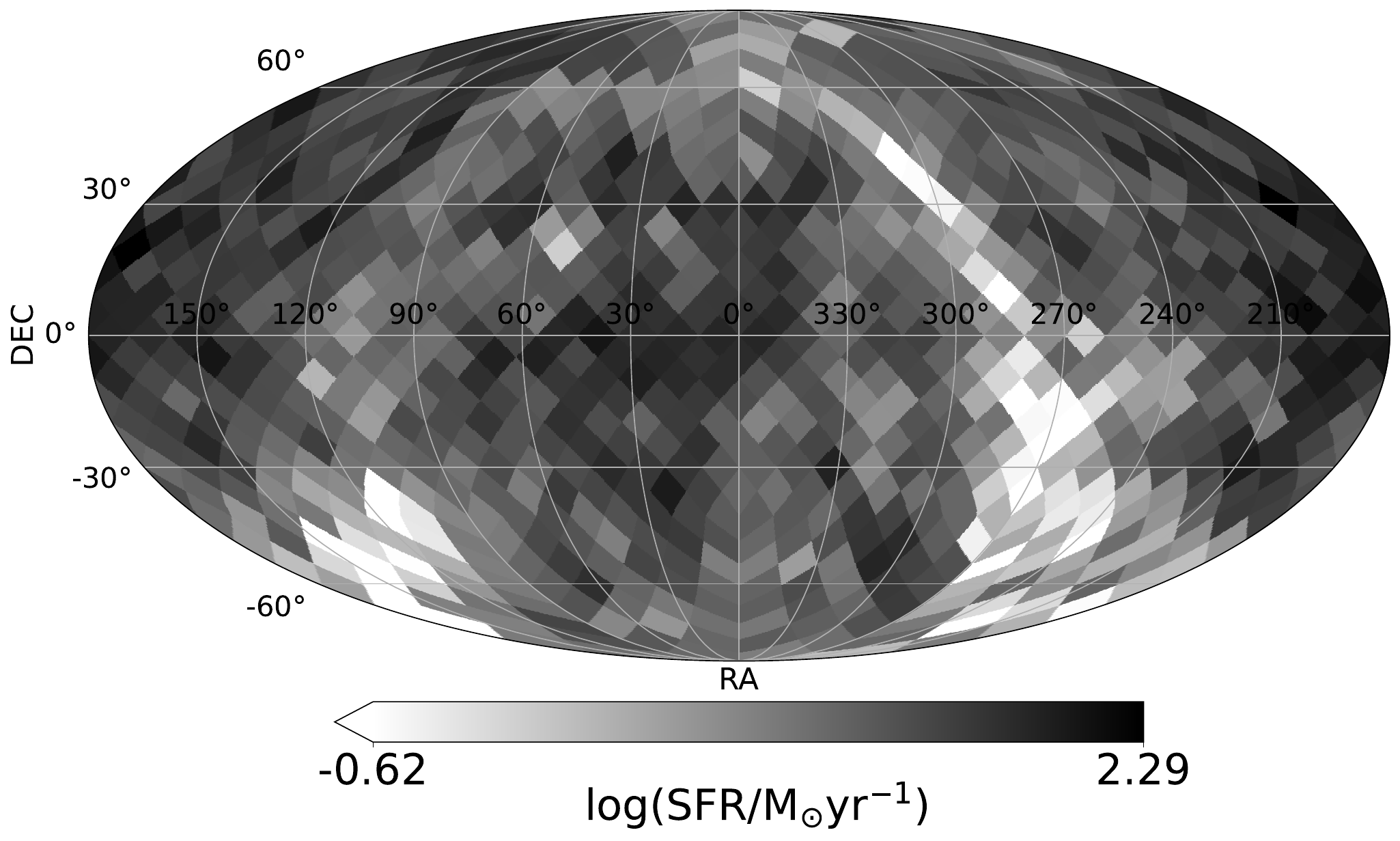}
\end{minipage}
}
\subfigure{
\begin{minipage}[b]{.3\linewidth}
\centering
\includegraphics[scale=0.16]{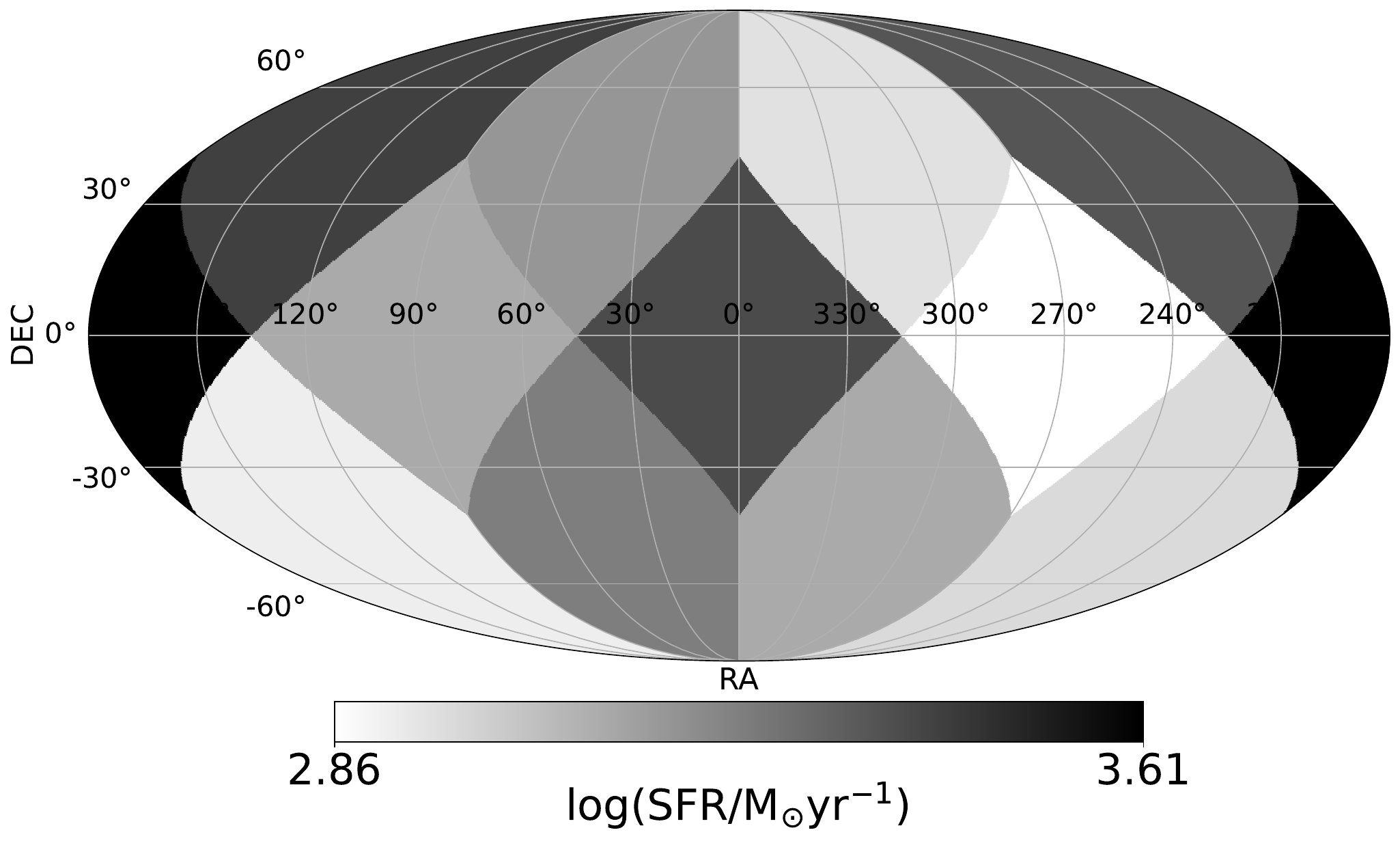}
\end{minipage}
}
\subfigure{
\begin{minipage}[b]{.3\linewidth}
\centering
\includegraphics[scale=0.16]{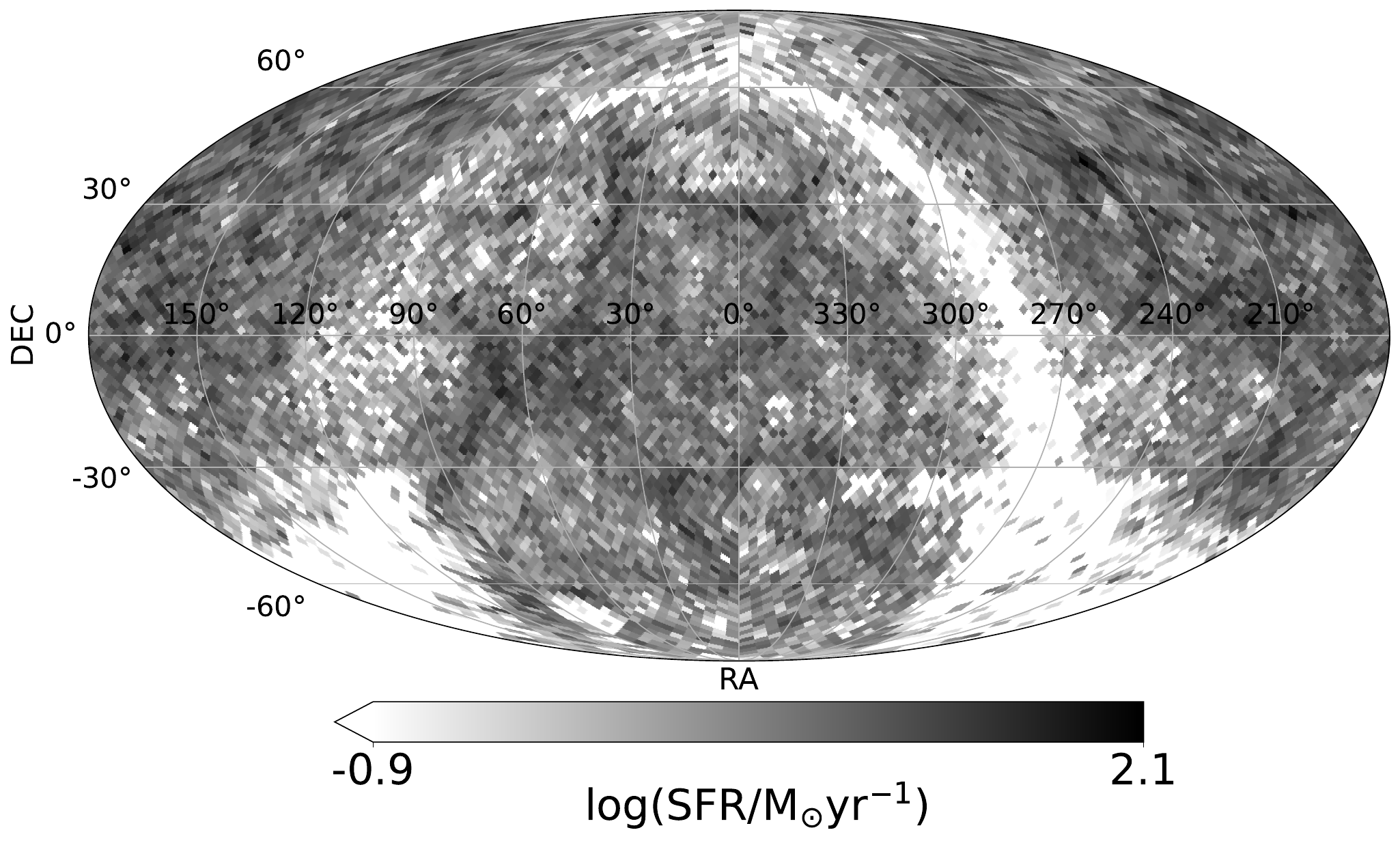}
\end{minipage}
}
\subfigure{
\begin{minipage}[b]{.3\linewidth}
\centering
\includegraphics[scale=0.16]{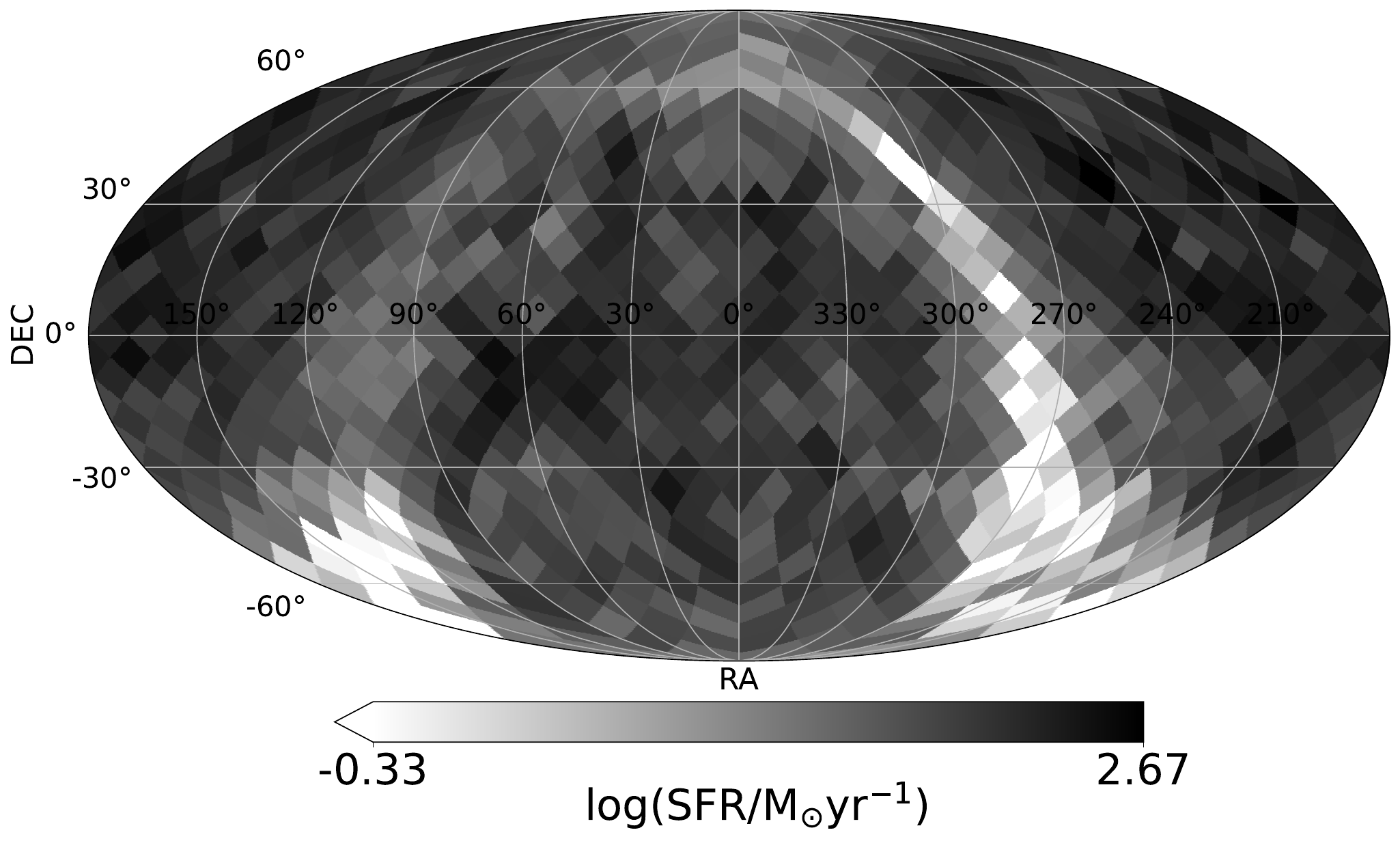}
\end{minipage}
}
\subfigure{
\begin{minipage}[b]{.3\linewidth}
\centering
\includegraphics[scale=0.16]{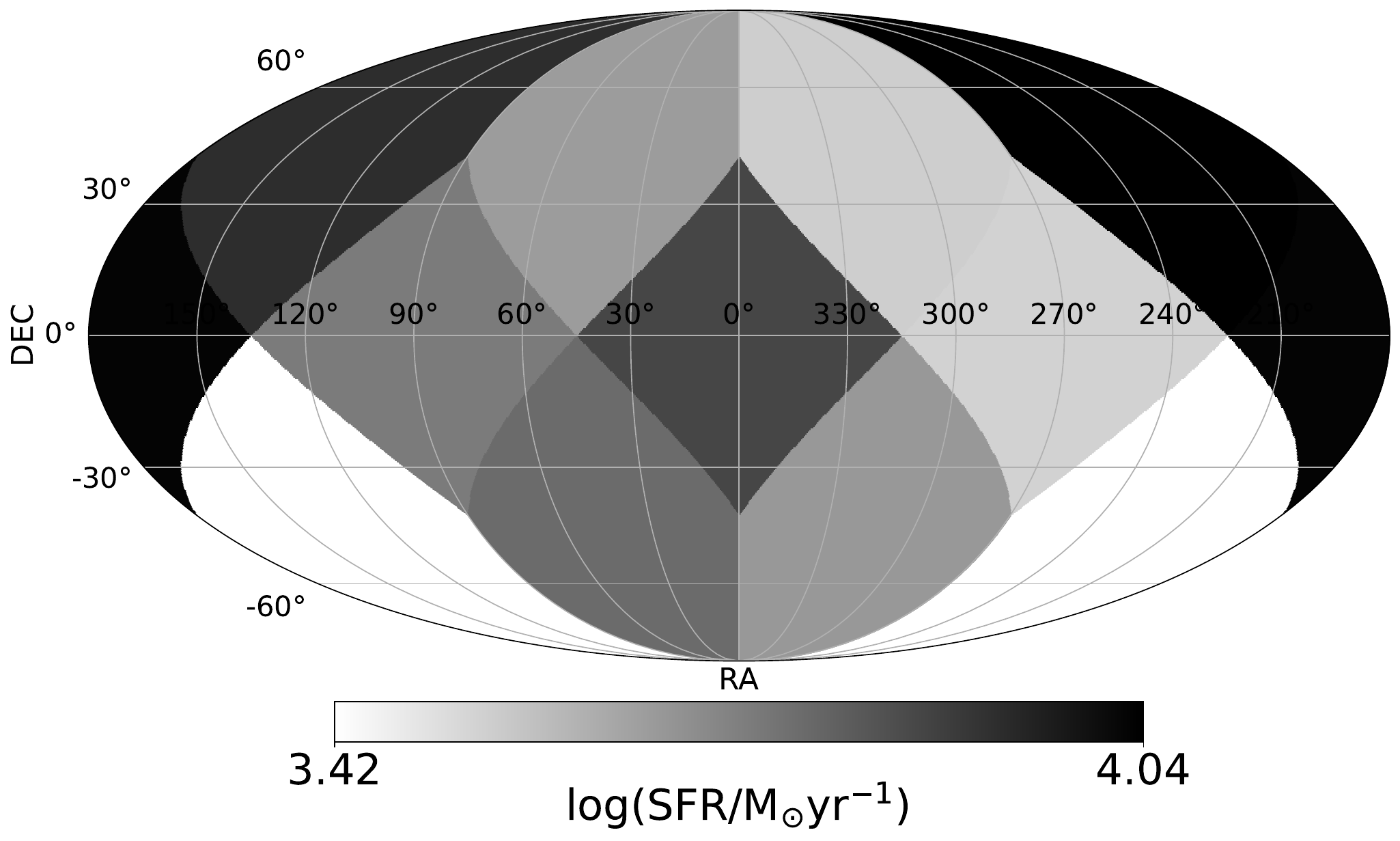}
\end{minipage}
}
\subfigure{
\begin{minipage}[b]{.3\linewidth}
\centering
\includegraphics[scale=0.16]{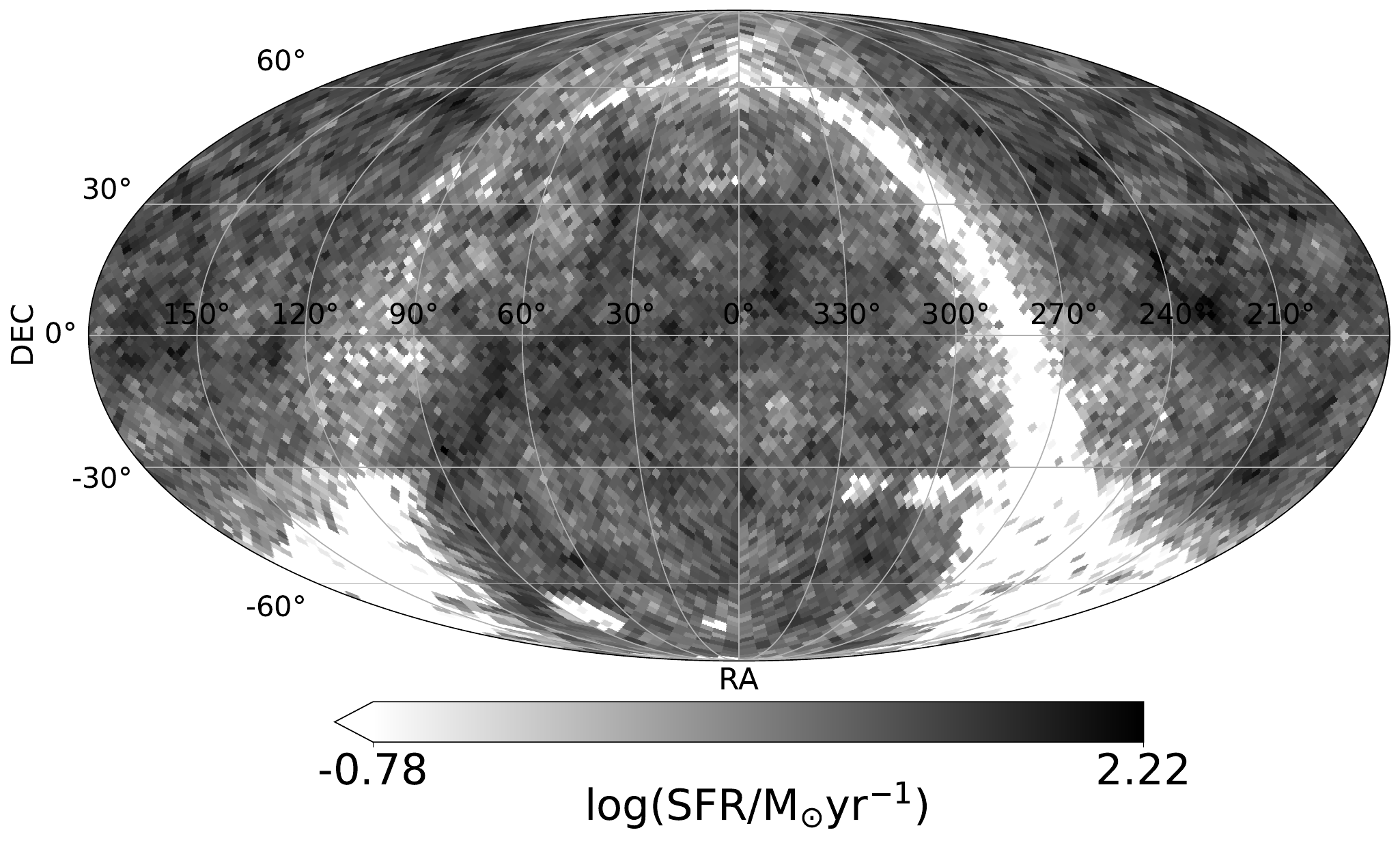}\\
\text{(a) $\theta=1.83^{\circ}$}
\end{minipage}
}
\subfigure{
\begin{minipage}[b]{.3\linewidth}
\centering
\includegraphics[scale=0.16]{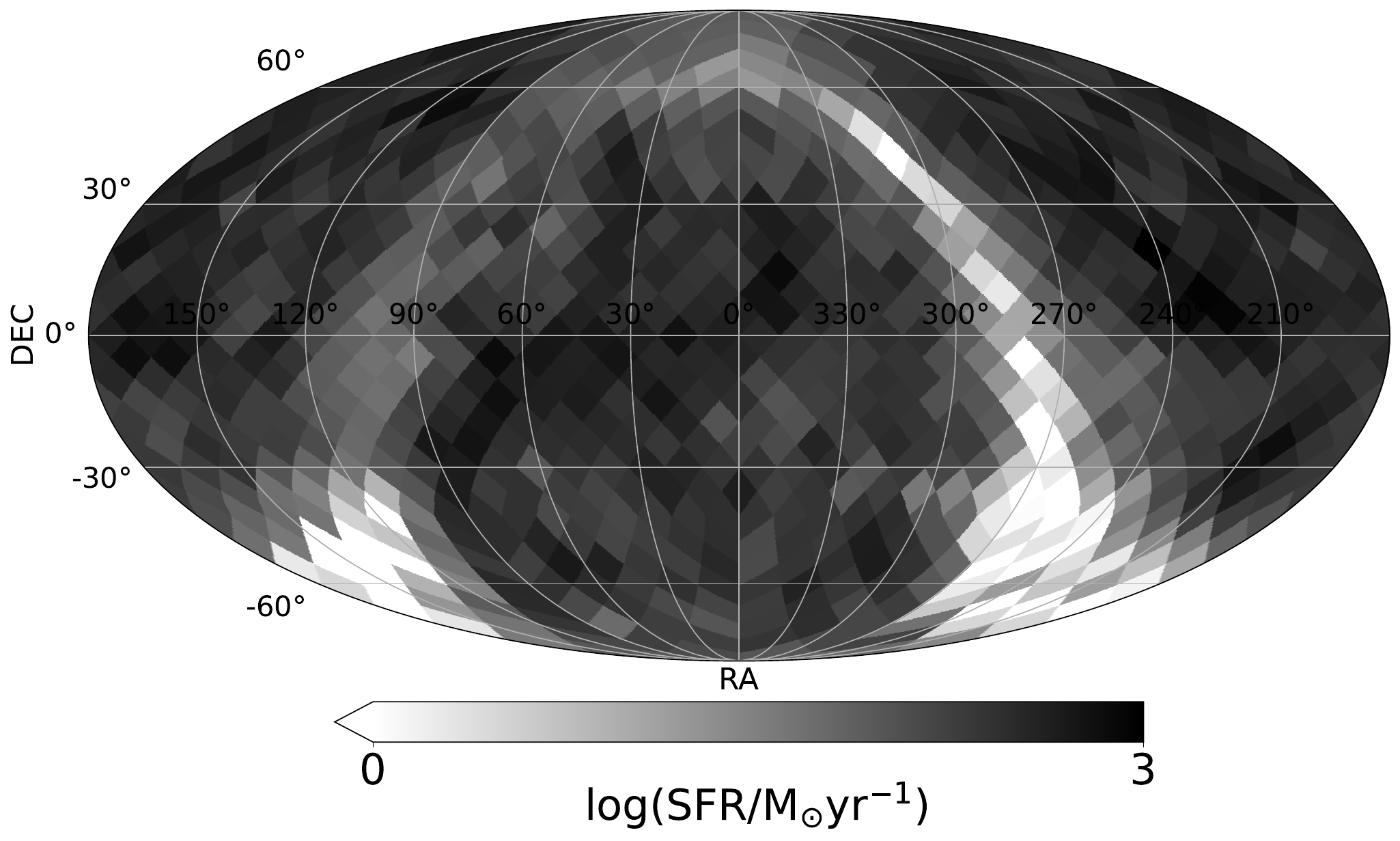}
\text{(b) $\theta=7.33^{\circ}$}
\end{minipage}
}
\subfigure{
\begin{minipage}[b]{.3\linewidth}
\centering
\includegraphics[scale=0.16]{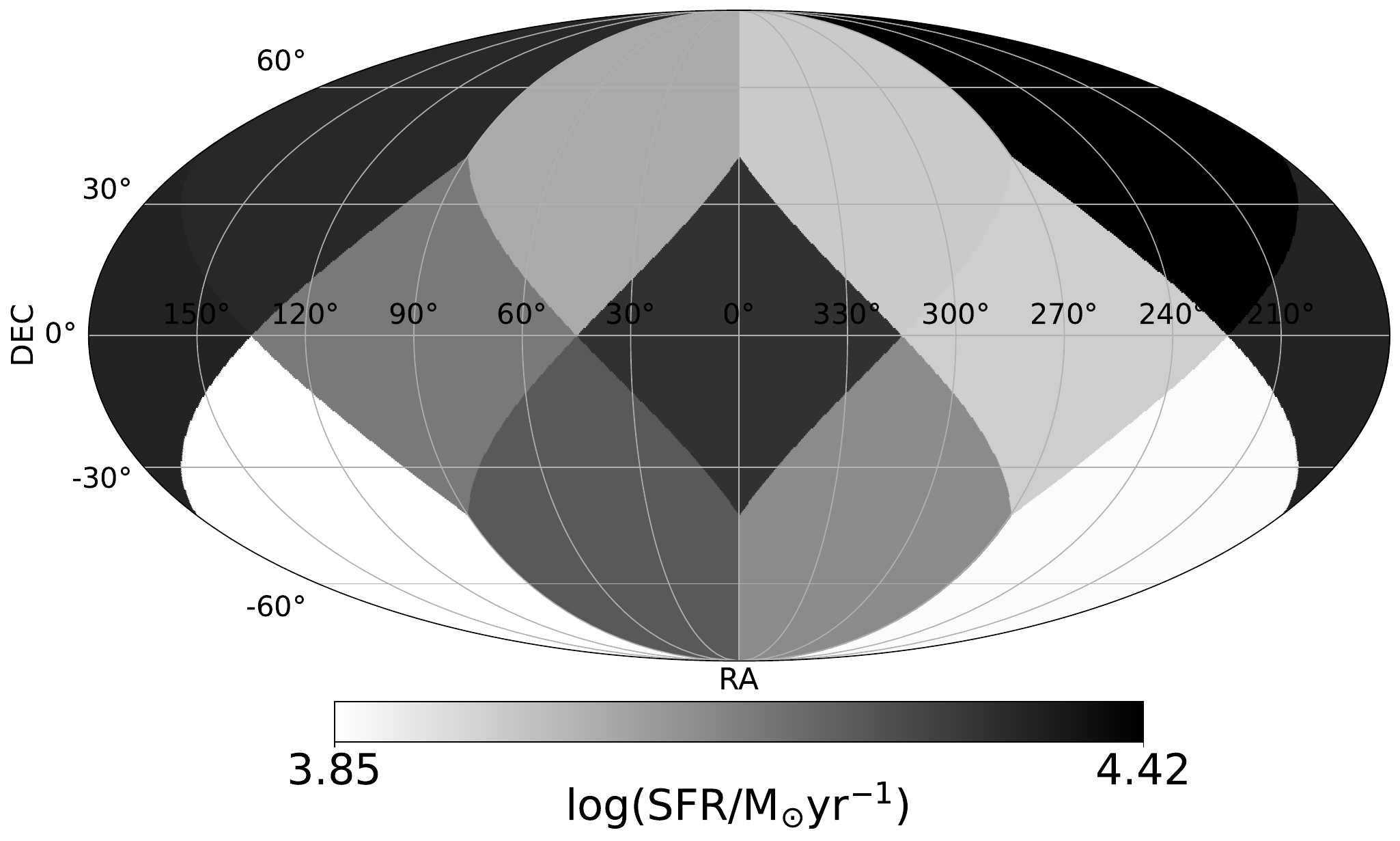}
\text{(c) $\theta=58.6^{\circ}$}
\end{minipage}
}
\caption{Sky distribution of SFR in equatorial coordinates. From left to right, the angular resolution of the grid of sky distributions at each column are $\theta=1.83^{\circ}$, $7.33^{\circ}$ and $58.6^{\circ}$, respectively. From top to bottom, the luminosity distance thresholds of sky distributions at each row are 50 Mpc, 100 Mpc, 150 Mpc, and 200 Mpc, respectively. The color bar range spans three orders of magnitude if the SFR range more than three orders of magnitude. Note: Due to the logarithmic scale of the color bar, the value 0 does not represent the actual SFR.}
\label{SFR_distribution}
\end{figure*}

The sky distributions of SFR of galaxies are shown in Fig.~\ref{SFR_distribution}, where column (a), (b) and (c) correspond to angular resolutions of the grids at $1.83^{\circ}$, $7.33^{\circ}$ and $58.6^{\circ}$, respectively. From top to bottom, the luminosity distance thresholds of the sky distributions of SFR in each row are 50 Mpc, 100 Mpc, 150 Mpc, and 200 Mpc, respectively. We can see that the sky distributions of SFR of galaxies exhibit patterns similar to the sky distributions of stellar mass of galaxies, particularly in terms of their main anisotropic structures. 
This similarity is expected, since massive galaxies typically have both greater stellar masses and higher SFRs.

The sky distributions of stellar mass and SFR exhibit pronounced variations under different FoVs and luminosity distance thresholds.
For example, in the stellar mass sky distribution within a luminosity distance threshold of 50 Mpc, the Virgo cluster contributes prominently at an angular resolution of $1.83^{\circ}$.
However, its contribution decreases substantially as the resolution becomes coarser: at $7.33^{\circ}$, the influence of the Virgo cluster is already greatly diminished.
Similar behavior is also observed in the sky distributions of SFRs.

\section{Discussions}
\label{sec:Discussion}

\subsection{Sample completeness issue of dwarf galaxies}
As the luminosity distance threshold increases, the observed galaxy population becomes increasingly dominated by massive systems, while faint dwarf galaxies gradually fall below the detection limit. This observational bias naturally prompts the question of how much dwarf galaxies truly contribute.
When filtering the data, we set the minimum quality limit at $10^5 \rm M_{\odot}$, excluding all galaxies below this threshold from their stellar mass estimates. \cite{2012MNRAS.421..621B} presented a well fitted galaxy stellar mass function (GSMF) in the nearby universe by
\begin{align}
\Phi(M)dM&=\left[\phi_1^{\ast} \left(\frac{M}{M_{\ast}} \right)^{\beta_1}+ \phi_2^{\ast} \left(\frac{M}{M_{\ast}} \right)^{\beta_2} \right]\nonumber\\
&\times\exp{\left(-\frac{M}{M_{\ast}} \right)} \frac{dM}{M_{\ast}}~,
\label{GSMF}
\end{align}
where $\Phi(M)dM$ is the number density of galaxies in the mass range of $M$ to $M+dM$, $\phi_1^{\ast}=2.93\times10^{-3}~\rm Mpc^{-3}$ and $\phi_2^{\ast}=0.63\times 10^{-3}~ \rm Mpc^{-3}$ are the normalized density, $M_{\ast}=\rm 10^{10.66}~M_{\odot}$ is the break mass, $\beta_1=-0.62$ and $\beta_2=-1.5$ are the power law indices \citep{2017MNRAS.470..283W}. We assume that the distribution of dwarf galaxies still follow the GSMF in Eq.(\ref{GSMF}). 
Then the fraction of unobservable dwarf galaxies is given by
\begin{equation}
f\lesssim \frac{\int^{\infty}_{\rm 10^5M_{\odot}}\Phi(M)MdM-\int^{\infty}_{\rm 10^8M_{\odot}}\Phi(M)MdM}{\int^{\infty}_{\rm 10^5M_{\odot}}\Phi(M)MdM}=1.78~\% ~.
\label{GSMF_integral}
\end{equation}
Here, we roughly assume that the minimum mass of the galaxies at the luminosity distance threshold is about $\sim10^8M_\odot$.
It is clear that dwarf galaxies ($10^5-10^8 \rm M_{\odot}$) constitute only a small fraction of the galaxy population. 
Therefore, the incompleteness of dwarf galaxy samples does not significantly affect the sky distributions of stellar mass and SFR that we derive.

\begin{figure*}
    \centering 
    \subfigure[Stellar mass ($\theta=1.83^{\circ}$)]{\label{subfig:a}
        \includegraphics[width=3.4in]{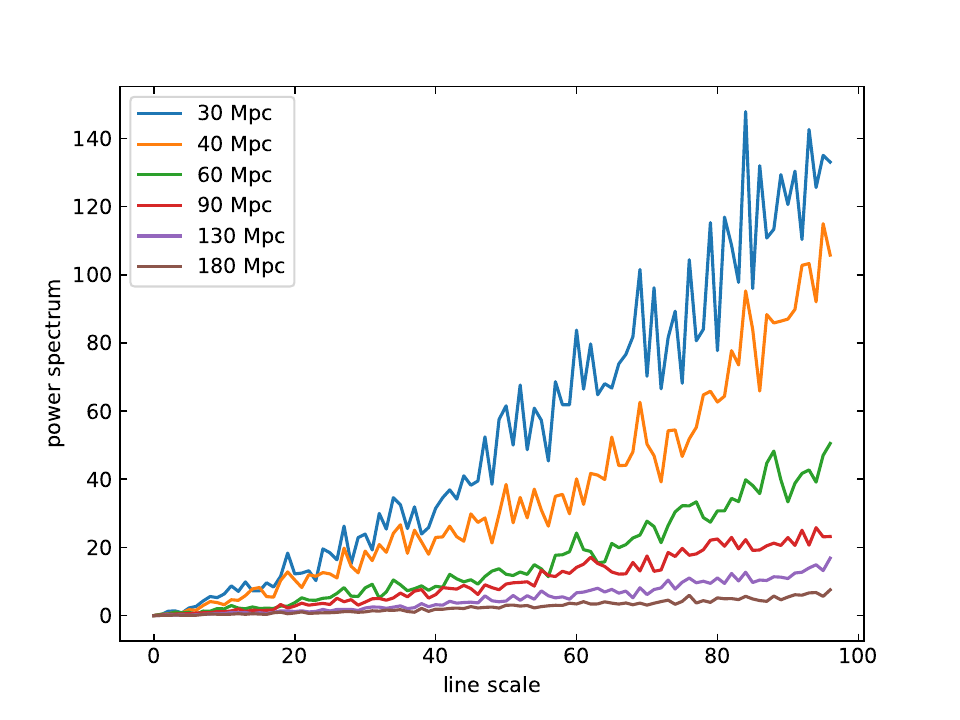}
    }
    \subfigure[SFR ($\theta=1.83^{\circ}$)]{\label{subfig:b}
        \includegraphics[width=3.4in]{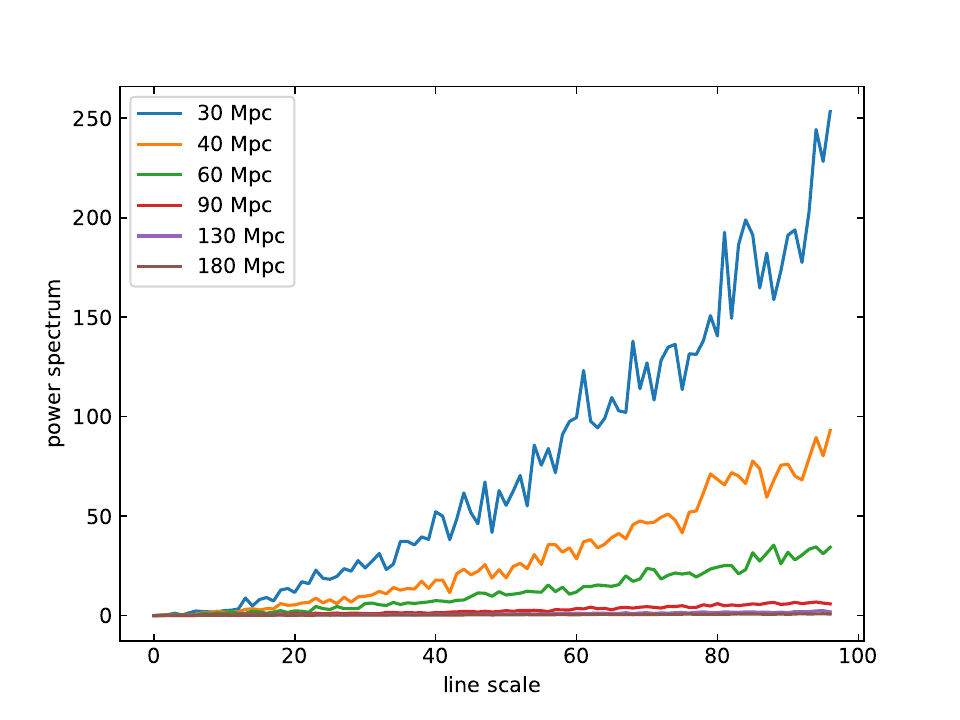}
    }  
    \\
    \subfigure[Stellar mass ($\theta=55^{'}$)]{\label{subfig:c}
        \includegraphics[width=3.4in]{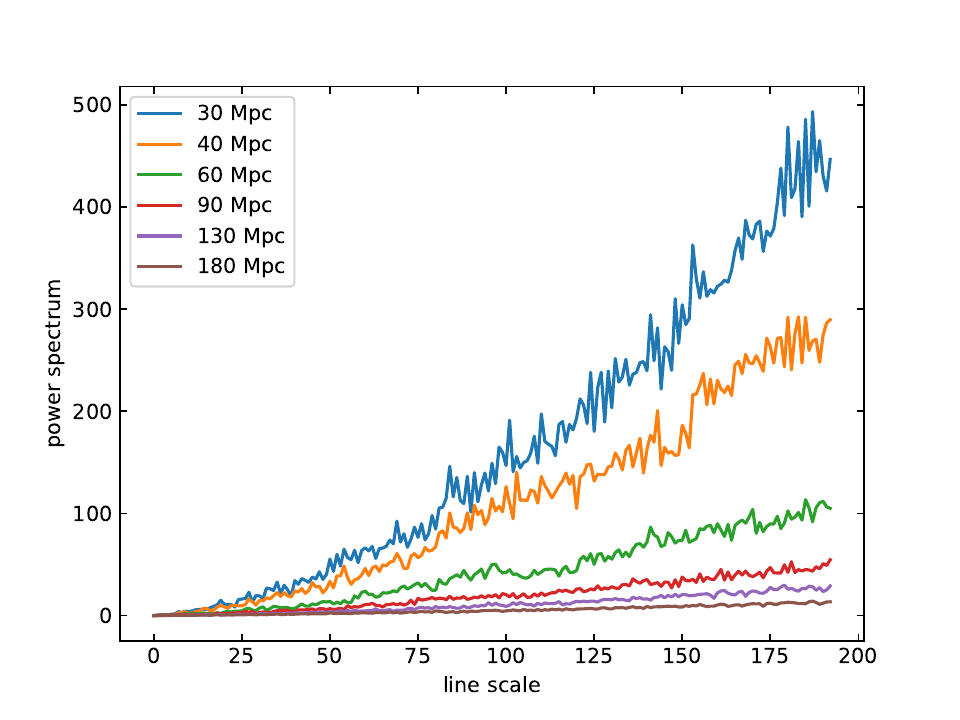}
    }
    \subfigure[SFR ($\theta=55^{'}$)]{\label{subfig:d}
        \includegraphics[width=3.4in]{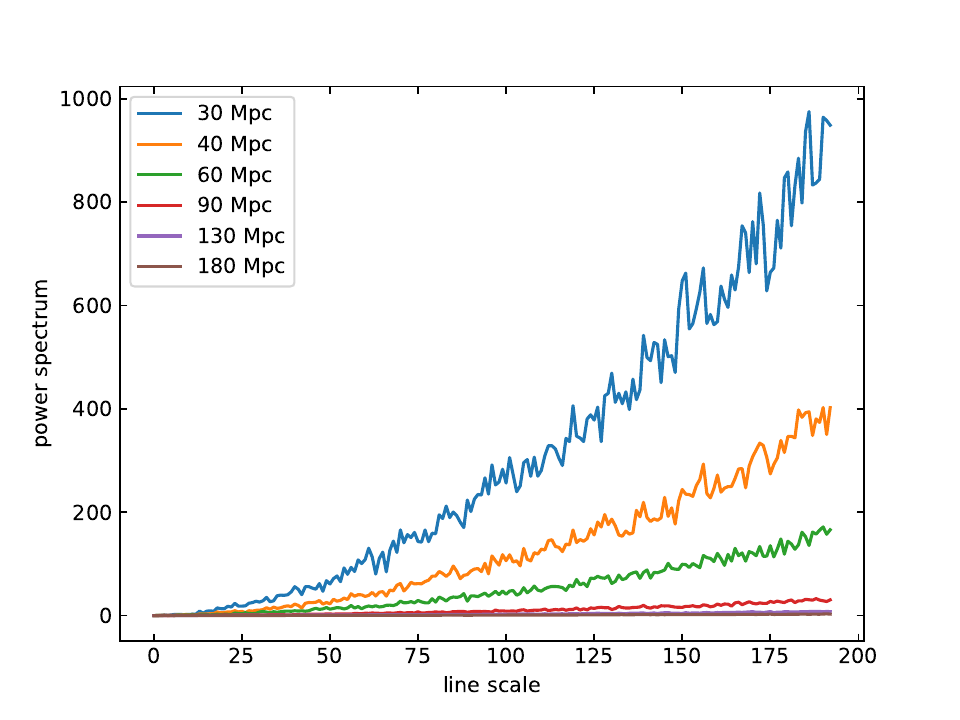}
    }

    \caption{Angular power spectra of the sky distributions of stellar mass (left columns) and SFR (right columns) of galaxies in the resolution $\theta = 1.83^{\rm \circ}$ (top rows) and $\theta=55^{'}$ (bottom rows).}
    \label{power_spectrum}
\end{figure*}

\setlength{\tabcolsep}{12pt}
\begin{deluxetable*}{lcccccccc}
  \tablecaption{The basic parameters for determining the size and quantity of the grid divisions. \label{theta_pix}}
  \tablecolumns{9}
  
  \startdata \\
    $N_{\rm side}$ & $1$ & $2$ & $4$ & $8$ & $16$ & $32$ & $64$ & $128$ \\
    \hline
    $\theta$ & $58.6^{\rm \circ}$ & $29.3^{\rm \circ}$ & $14.7^{\rm \circ}$ & $7.33^{\rm \circ}$ & 
    $3.66^{\rm \circ}$ & $1.83^{\rm \circ}$ & $55.0^{'}$ & $27.5^{'}$ \\
     $N_{\rm grid}$& $12$ & $48$ & $192$ & $768$ & $3072$ & $12288$ & $49152$ & $196608$ \\
\enddata
\tablecomments{$N_{\rm side}$ is the number of divisions along each side of
a base grid. $\theta$ is the angular resolution of the grid and $N_{\rm grid}$ is the number of grids \citep{healpix}.}
\end{deluxetable*}
\setlength{\tabcolsep}{6pt}

\subsection{Anisotropic structures of sky distributions}
As shown in Fig.~\ref{mass_distribution} and Fig.~\ref{SFR_distribution}, the sky distributions of stellar mass and SFR show an increasingly isotropic trend as the luminosity distance threshold increases. To characterize the anisotropic structures of the sky distributions, we employ the angular power spectrum, a method commonly used to analyze anisotropies in the Cosmic Microwave Background \citep{2007ApJSpowerspectrum}. We utilized the two-point correlation function to calculate the angular power spectra of the stellar mass and SFR sky distributions.

We first masked the Galactic plane ($-10^{\circ}$ to $10^{\circ}$) due to limited observational data caused by dust extinction in the Milky Way. 
For the unmasked region, we first calculated the mean SFR/stellar mass. We then determined the deviation of each grid from this mean and normalized it by dividing by the mean itself, yielding a dimensionless quantity. According to the \texttt{HEALPix} grid division scheme, the celestial sphere is initially partitioned into twelve base pixels. The resolution of the grid is defined by \( N_{\rm side} \), which represents the number of divisions along each side of a base pixel \citep{healpix}. Consequently, the total number of pixels is \( N_{\rm grid} = 12\,N_{\rm side}^2 \), and the approximate angular resolution grid is given by \citep{healpix} 
\begin{equation}
   \theta\sim\sqrt{\frac{4\pi}{12N^2_{\rm side}}}=\sqrt{\frac{\pi}{3N^2_{\rm side}}}.
   \label{eq:angular resolution}
\end{equation}
Therefore, we estimate the maximum multipole moment by
\begin{equation}
    \ell_{\rm max}\sim \frac{\pi}{\theta}=\pi \sqrt{\frac{3N^2_{\rm side}}{\pi}} \approx3N_{\rm side}.
\end{equation}
We compute the angular power spectra for the sky distributions with an angular resolution of $1.83^{\circ}$ and $55^{'}$, which corresponds to $N_{\rm side}=32$ and 64 (see Table ~\ref{theta_pix}). Fig.~\ref{power_spectrum} presents the angular power spectra for the SFR and stellar mass sky distributions at luminosity distance thresholds of 30, 40, 60, 90, 130, and 180 Mpc.
We summarized the following properties of the angular power spectra: (i) The angular power spectra are stronger at nearly distances, indicating that the stellar mass/SFR distribution is more inhomogeneous at closer distance, with enhanced structural features. As distance increases, the angular power spectra drops rapidly toward zero, reflecting a trend toward homogeneity in the distribution. (ii) At a given luminosity distance, higher angular resolution (i.e., smaller grid scales) strengthens the angular power spectra. This occurs because high angular resolution increase the number of empty grid cells, amplifying inhomogeneity and structural complexity in the stellar mass/SFR distribution. (iii) At given angular resolution, the relative strength of the angular power spectra of stellar mass and SFR varies with distance. For example, the angular power spectra of SFR decreases more rapidly from 30 Mpc to 40 Mpc compared to that of stellar mass. 
This indicates that stellar mass and SFR exhibit different dependencies on distance in their respective distributions.
Such variations may also reflect differences in the spatial distributions of various transient sources. (iv) At a given angular resolution, the angular power spectrum of SFR is initially stronger than that of stellar mass (e.g., 30 Mpc) but decays toward zero more rapidly. 
The definition of fluctuations relative to the average value is $\Delta=(X_i-\bar{X})/\bar{X}$, where $X_i$ is total stellar mass/SFR of each grid and $\bar{X}$ represents the average value of all grids excluding the plane of the Milky Way ($\pm 10^{\rm \circ}$), which yield $\Delta=-1$ in the empty grids. As a result, when non-empty grids contain very high stellar mass values, the extreme contrast between non-empty and empty grids generates significant angular power spectra at high $\ell$. This explains why the angular power spectrum exhibits a monotonic increase with $\ell$.
Furthermore, the angular power spectra shows a systematic weakening of the two-point correlation with increasing luminosity distance threshold, indicating a transition from anisotropic to more isotropic sky distributions. 
This trend is reflected in the angular power spectrum approaching zero at 180 Mpc, indicating that the distributions become highly isotropic at these large distances.

\begin{figure}
 \includegraphics[width=\columnwidth]{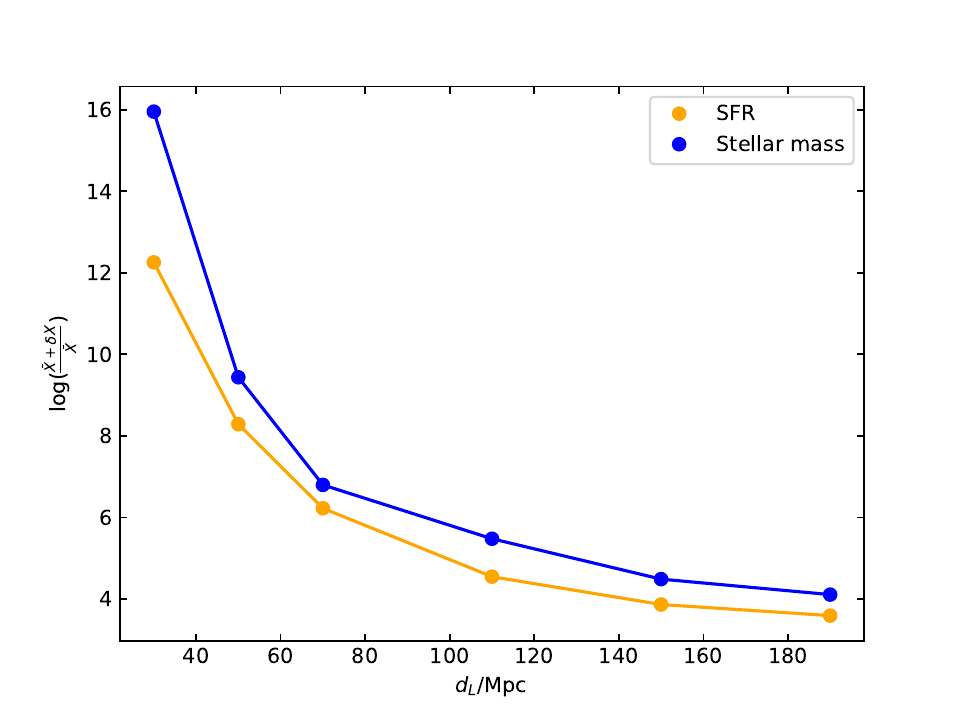}
\caption{Relative fluctuations in the sky distributions of stellar mass and SFR at different luminosity distance thresholds (30 Mpc, 50 Mpc, 70 Mpc, 110 Mpc, 150 Mpc, 190 Mpc).
\label{fluctuation}}
\end{figure}

In addition to the angular power spectra analysis, we also calculated the relative fluctuation in the distributions of stellar mass and SFR using $(\bar{X}+\delta X)/\bar{X}$, where $\delta X$ represents the root mean square of the grid values. Fig.~\ref{fluctuation} presents the relative fluctuations in the distributions at luminosity distance threshold of 30, 50, 70, 110, 150, and 190 Mpc. 
As the luminosity distance threshold increases, the amplitude of fluctuations decreases, also reflecting a gradual progression toward a more isotropic and homogeneous universe.

\subsection{Application to CCSNe distributions}
\begin{figure*}[h]
\begin{center}
\subfigure{
    \includegraphics[width=3.3in,height=2.1in]{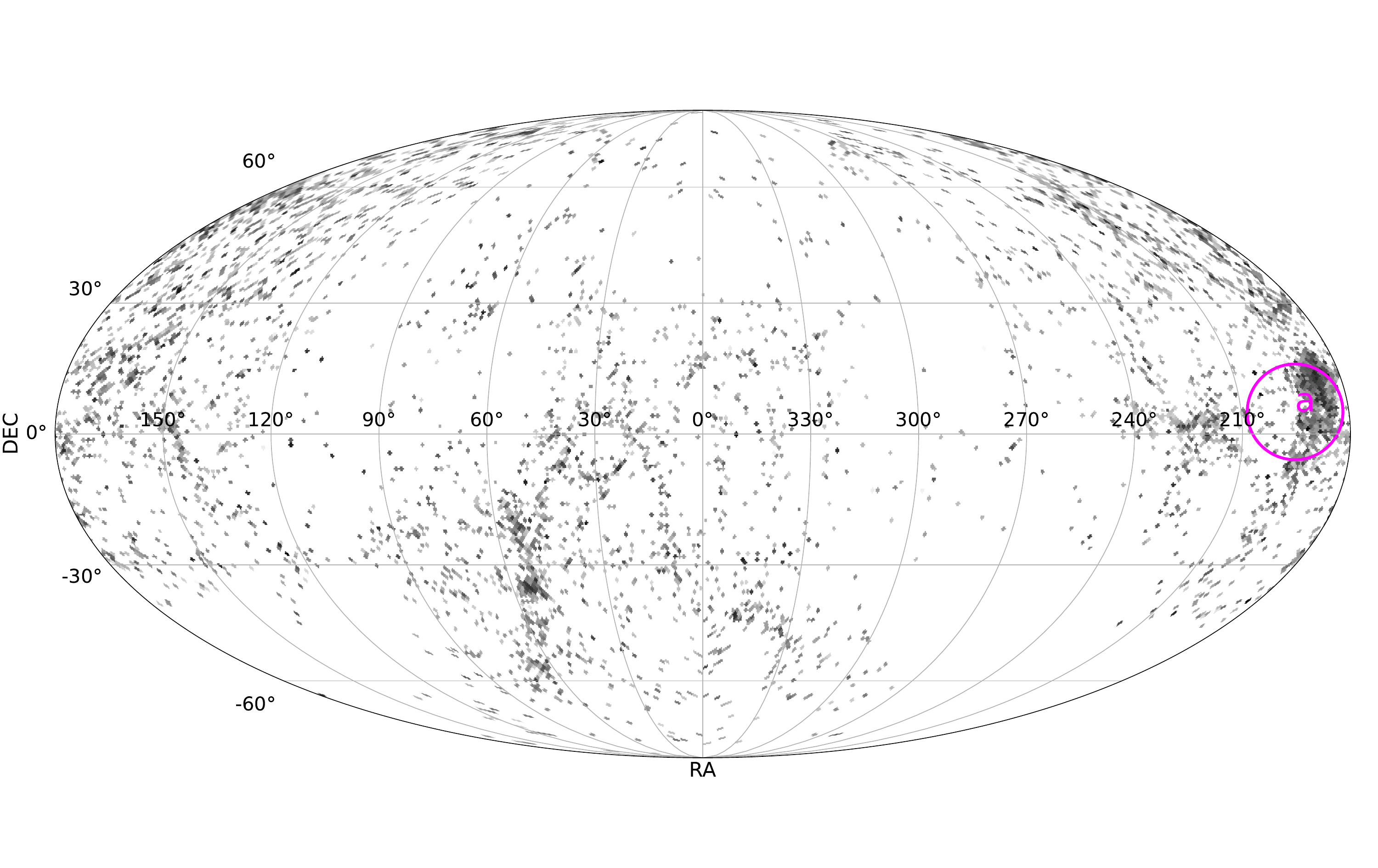}
}
\subfigure{
    \includegraphics[width=3.3in,height=2.1in]{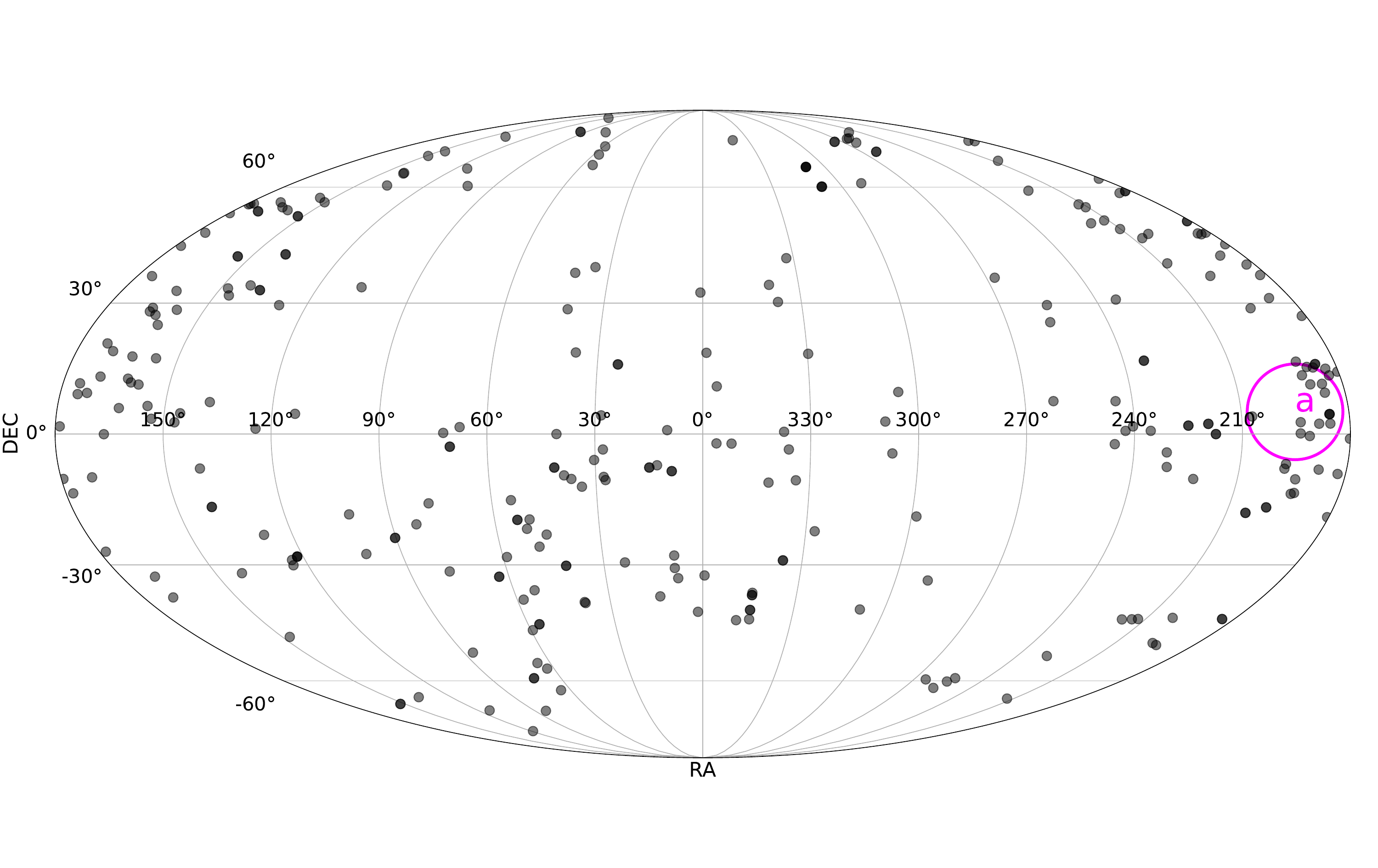}
}  
\subfigure{
    \includegraphics[width=3.3in,height=2.1in]{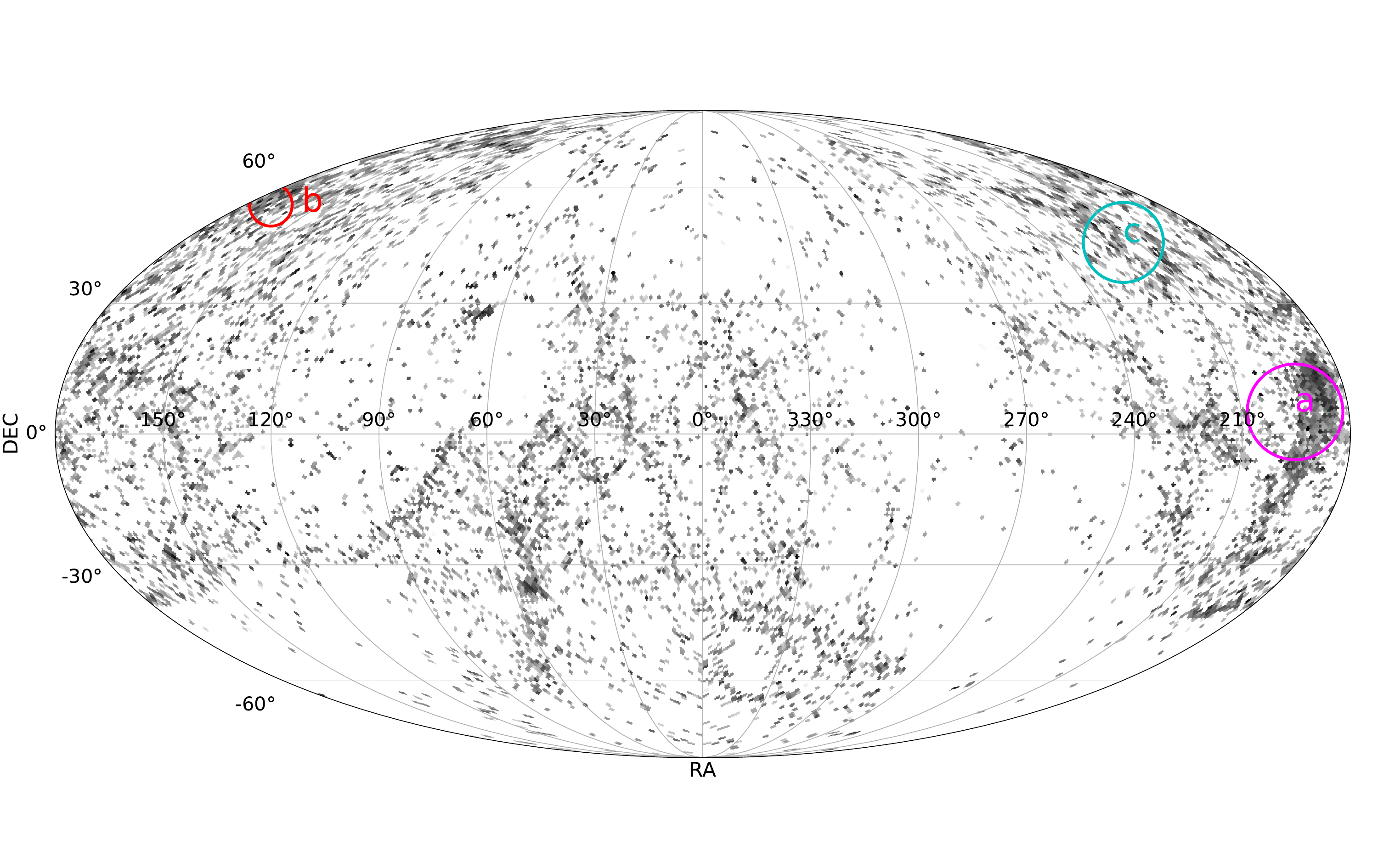}
}
\subfigure{
    \includegraphics[width=3.3in,height=2.1in]{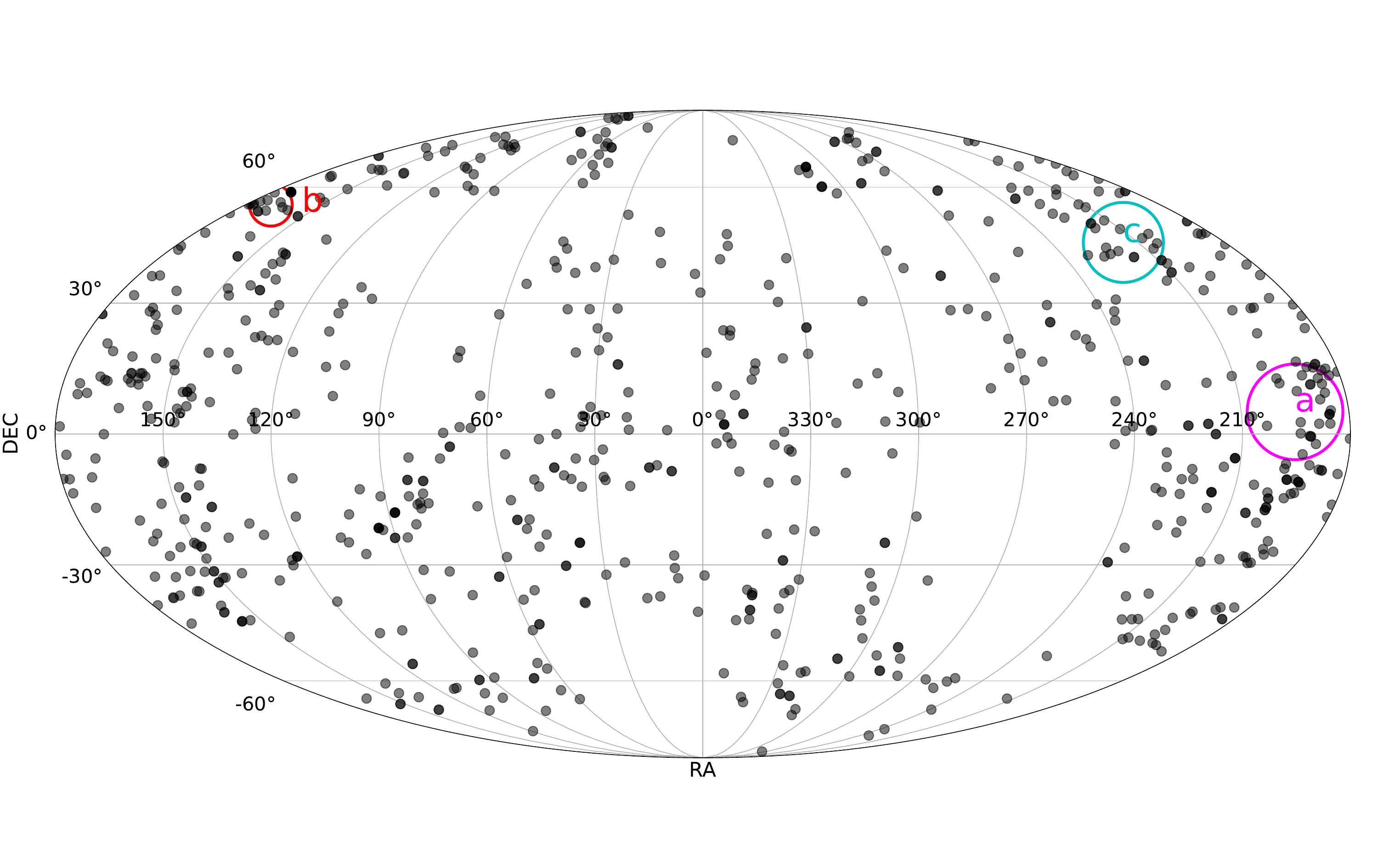}
}
\subfigure{
    \includegraphics[width=3.3in,height=2.1in]{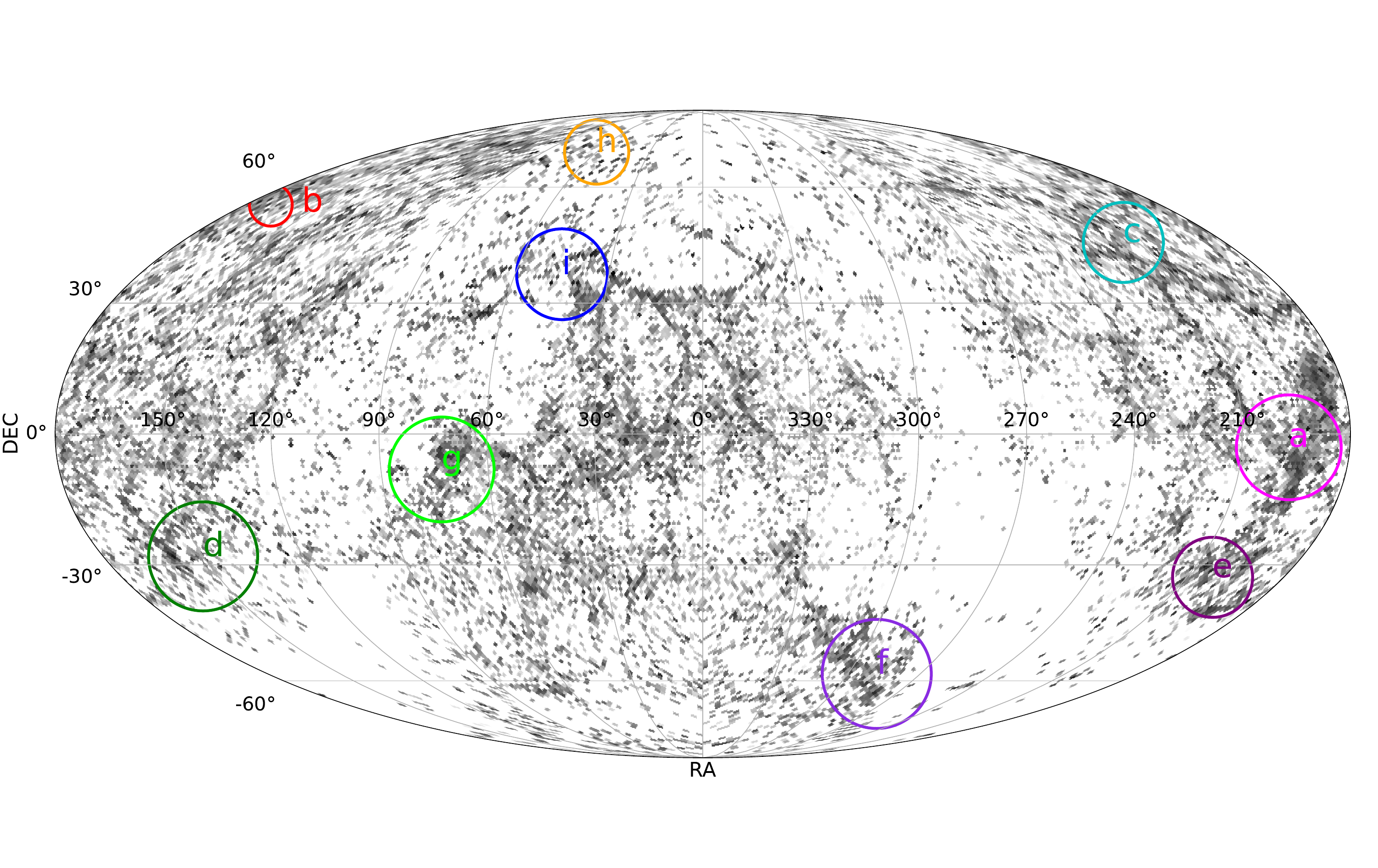}
}
\subfigure{
    \includegraphics[width=3.3in,height=2.1in]{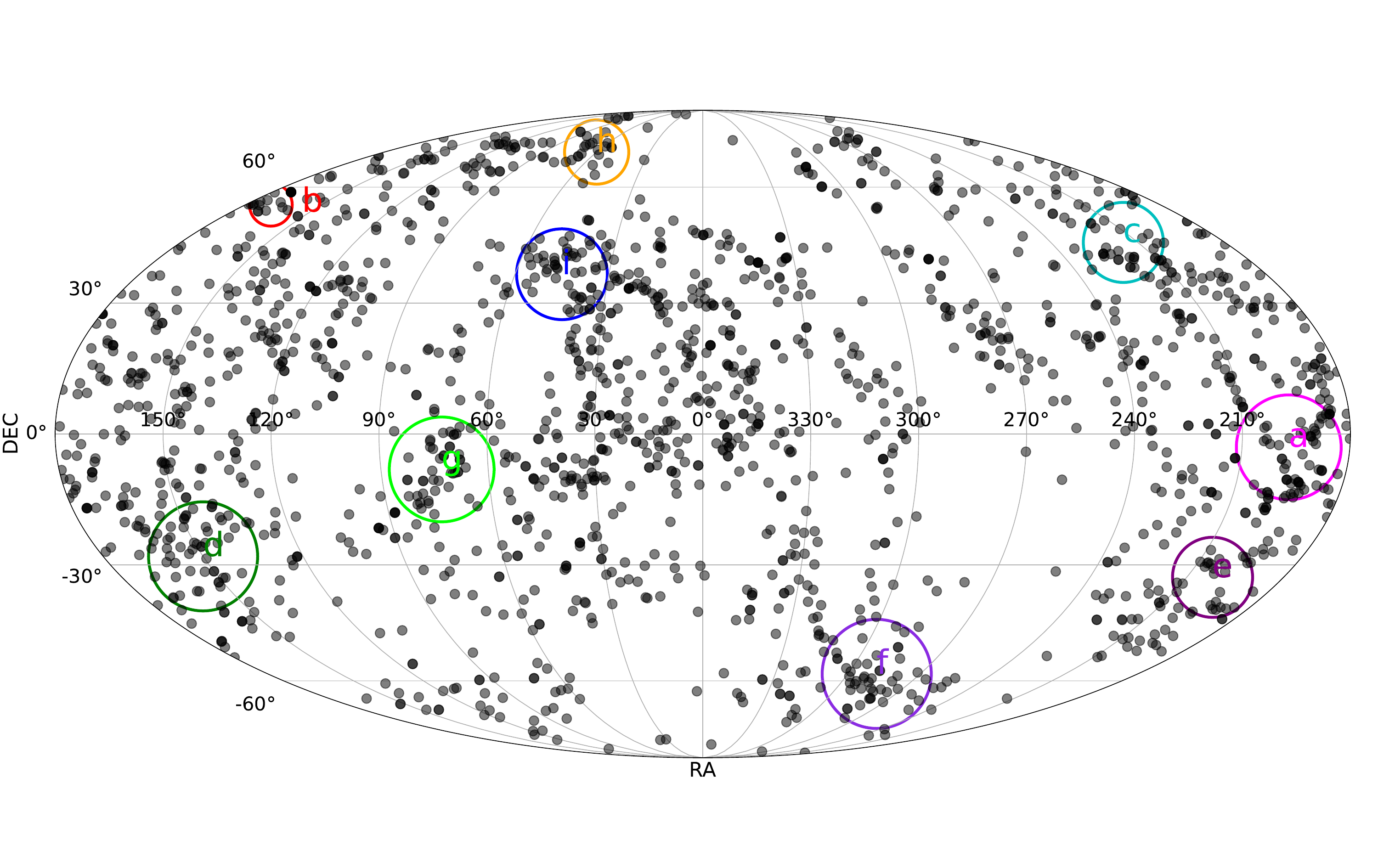}
}

\caption{
Comparison of the sky distribution of SFR with that of CCSNe. The left column displays the sky distributions of SFR of galaxies at an angular resolution of the grid $\theta=55^{'}$. The right column shows scatter plots of CCSNe. From top to bottom, the luminosity distance thresholds are 30 Mpc, 50 Mpc, and 80 Mpc, respectively. All the color bars of SFR distributions span three orders of magnitude but are not displayed. All the figures are based on equatorial coordinates. Each row represents a comparison at the same luminosity distance threshold, and each colored circle corresponds to each other.}

\label{Matching}
\end{center}
\end{figure*}

The provided sky distributions of stellar mass and SFR are intended to serve as a guide for designing optimized sky surveys aimed at discovering transients. 
In this section, we perform a qualitative comparison between the sky distributions of SFR and CCSNe. We collect supernovae data from the Bright Supernova catalog \footnote{{\href{https://www.rochesterastronomy.org/snimages/}{https://www.rochesterastronomy.org/snimages/}}} (BSN; \citealt{BrightSN2013}) and  Transient Name Server (TNS)\footnote{{\href{https://www.wis-tns.org/}{https://www.wis-tns.org/}}}  to match the sky distributions of SFR of galaxies. The BSN and TNS contain more than 51,000 and 18,000 supernovae with redshifts, respectively. After excluding Type Ia supernovae, which are not from the core collapse of massive stars, subsets of approximately 7,222 and 5,682 supernovae remain in the BSN and TNS, respectively, which have catalogued redshifts, coordinates, and type information. We analyzed the spatial distribution of supernovae using their celestial coordinates and correlated this distribution with the SFR sky distribution at an angular resolution of $\theta=55^{'}$. 
Fig.~\ref{Matching} presents a comparison of these two distributions at luminosity distance thresholds of 30, 50, and 80 Mpc, which contain 302, 761, and 1803 supernovae, respectively. In Fig.~\ref{Matching}, prominent regions of enhanced SFR are marked with colored circles to denote areas with a high concentration of supernova discoveries. Within this SNe sample, the overlap between the BSN and TNS catalogs includes 132, 370, and 1,009 SNe at distances of 30, 50, and 80 Mpc, respectively. At these same distance thresholds, 144, 342, and 694 SNe are unique to the BSN, while 26, 49, and 100 SNe are unique to the TNS.
At larger luminosity distance thresholds, prominent structures in the sky distribution emerge in the SFR sky distributions, which is consistent with the relationship between CCSNe and SFR \citep{1998MNRAS.297L..17M,2012ARA&A..50..531K}. Consequently, these SFR and stellar mass sky distributions can guide and optimize survey strategies for detecting related extragalactic transients.

However, we need to emphasize that the comparison presented above is strictly qualitative in nature. The current supernova samples compiled from the BSN and TNS are a complex mixture of discoveries from both historical targeted surveys (e.g., LOSS, CHASE) and modern untargeted wide-field surveys (e.g., ATLAS, ZTF, LSST). These surveys differ substantially in their target selection strategies, exposure times, cadences, and sky coverage, all of which introduce significant and complex observational biases into the observed spatial distribution of SNe. For example, targeted surveys preferentially monitored luminous, massive galaxies, leading to an overrepresentation of SNe in high-mass hosts, while untargeted surveys systematically scan the sky regardless of galaxy properties, yielding a more complete but still survey-dependent snapshot. Deconvolving these intertwined selection effects to enable a rigorous, quantitative comparison between observed SN counts and our intrinsic SFR maps is currently unfeasible and would risk producing misleading results. Consequently, we restrict our analysis to a qualitative spatial comparison. Our modest objective is to verify whether the prominent large-scale structures (e.g., the Virgo cluster) in our SFR maps broadly align with the historical spatial footprints of CCSNe.

Regarding the time interval of the SN sample, because our qualitative comparison aims only to trace time-integrated, static spatial structures rather than to calculate temporal event rates or volumetric rates, we include all SNe with available redshift, coordinates, and type information across all available time periods to maximize the statistical sample size for spatial alignment. We emphasize that unclassified transients and Type Ia supernovae are strictly excluded from this validation sample to ensure purity in testing the spatial correlation between confirmed CCSNe and current SFR. Despite these limitations, the broad consistency between the overdensities in our SFR maps and the spatial clustering of historical CCSNe provides encouraging support for using these maps as intrinsic ``prior probability maps" for future transient surveys. Once future surveys with well-characterized selection functions and uniform cadences become available, our maps can serve as a clean baseline onto which these observational selection effects can be explicitly convolved for rigorous quantitative studies.

\section{Conclusion}
\label{sec:Conclusion}
Different transients originate from distinct progenitors associated with either stellar mass or SFR. In the nearby universe, both of these quantities are approximately isotropic. To enhance the efficiency of transient detection in surveys, we calculate the sky distributions of stellar mass and SFR at various luminosity distance thresholds, using a relatively complete sample of nearby galaxies.
These distributions can be used to characterize the anisotropic spatial distributions of nearby extragalactic transients and to inform the design of future telescope survey strategies.
These distributions exhibit anisotropic structures that gradually become more isotropic as the luminosity distance threshold increases. To maximize the likelihood of detecting transients, observational campaigns should prioritize these prominent over-dense regions, where galaxies are most highly concentrated.

We computed the angular power spectra of the sky distributions of stellar mass and SFR at the given angular resolutions $\theta=1.83^{\circ}$ and $55^{'}$. The analysis reveals that the sky distributions of stellar mass and SFR exhibit significant anisotropic structures at nearby distances but rapidly approach isotropy as the luminosity distance threshold increases. Furthermore, the differences in their angular power spectra suggest distinct spatial distributions and different event rates for associated transients.

Additionally, we quantified the amplitude of fluctuations in the stellar mass and SFR sky distributions across different luminosity distance thresholds. The observed decrease in these fluctuations with increasing distance provides evidence that the universe becomes progressively more isotropic on larger scales.
Consequently, the sky distributions we derived provide valuable guidance for planning future transient surveys.

Furthermore, our sky distribution maps have direct practical applications for current and near-term survey facilities. For example, the Mephisto telescope, a 1.6-meter multi-channel photometric survey telescope with a FoV of $3.14~\rm deg^2$, developed and operated by the South-Western Institute for Astronomy Research, Yunnan University — can simultaneously observe the same field in three channels (e.g., $ugi$, $vrz$) for transient searches and rapid identification. When designing its survey strategy, our stellar mass and SFR maps at the corresponding angular resolution ($\theta=1.83^{\circ}$) can be used to prioritize high-density regions, thereby significantly enhancing the probability and efficiency of discovering local transients. More generally, our method provides a flexible framework that can be adapted to the specific FoV of any telescope, generating customized sky distributions at different luminosity distance thresholds to inform and optimize survey strategies for various transient search programs.

The full dataset, machine learning training samples and related codes are available on Zenodo: \dataset[https://doi.org/10.5281/zenodo.20695048]{https://doi.org/10.5281/zenodo.20695048}. We provide various sky distributions of stellar mass and SFR, with angular resolutions ranging from $58.6^{\circ}$ to $27.5'$ (see Table ~\ref{theta_pix}) and luminosity distance thresholds coverage from 30 Mpc to 200 Mpc in steps of 10 Mpc.

\section{Acknowledgement}
We thank the anonymous referee for providing helpful comments and suggestions.
We acknowledge the helpful discussions with Jianhui Lian, Ziwei Li, Helong Guo, Gergely Dálya and Hugo Tranin. A part of numerical computations were conducted on the Yunnan University Astronomy Supercomputer. 
This research has made use of the VizieR catalogue access tool, CDS,
 Strasbourg, France (DOI : 10.26093/cds/vizier). The original description 
 of the VizieR service was published in 2000, A\&AS 143, 23. This research has made use of the NASA/IPAC Extragalactic Database (NED), which is operated by the Jet Propulsion Laboratory, California Institute of Technology, under contract with the National Aeronautics and Space Administration. This work has also utilized data from the NASA/IPAC Infrared Science Archive (IRSA), which is funded by the National Aeronautics and Space Administration and operated by the California Institute of Technology.
This work is supported by the National Key Research and Development Program of China (No. 2024YFA1611603, 2024YFA1611704), the National Natural Science Foundation of China (No. 12473047, 12393813), the Yunnan Key Laboratory of Survey Science (No. 202449CE340002), and the Strategic Priority Research Program of the Chinese Academy of Sciences (grant No. XDB0550400).

\bibliography{sample631}{}
\bibliographystyle{aasjournal}

\end{document}